\documentclass[%
 reprint,
 amsmath,amssymb,
 aps,prb, twocolumn, showpacs
]{revtex4-1}

\usepackage{graphicx}
\usepackage{dcolumn}
\usepackage{bm}
\usepackage{float}
\usepackage{subfigure}
\usepackage{amsmath}
\usepackage{mathrsfs}
\usepackage{amssymb}
\usepackage{hyperref}

\begin{document}

\preprint{}

\title{Poisson-Boltzmann Equation with a Random Field for Charged Fluids}

\author{Li Wan}
\email{lwan@wzu.edu.cn}
\affiliation{Department of Physics, Wenzhou University, Wenzhou 325035, P. R. China}
\author{Ning-Hua Tong}
\affiliation{Department of Physics, Renmin University of China, 100872 Beijing, P. R. China}
\date{\today}

\begin{abstract}
The classical Poisson-Boltzmann equation (CPBE), which is a mean field theory by averaging the ion fluctuation, has been widely used to study ion distributions in charged fluids. In this study, we derive a modified Poisson-Boltzmann equation with a random field from the field theory and recover the ion fluctuation through a multiplicative noise added in the CPBE. The Poisson-Boltzmann equation with a random field (RFPBE) captures the effect of the ion fluctuation and gives different ion distributions in the charged fluids compared to the CPBE. To solve the RFPBE, we propose a Monte Carlo method based on the path integral representation. Numerical results show that the effect of the ion fluctuation strengthens the ion diffusion into the domain and intends to distribute the ions in the fluid uniformly. The final ion distribution in the fluid is determined by the competition between the ion fluctuation and the electrostatic forces exerted by the boundaries. The RFPBE is general and feasible for high dimensional systems by taking the advantage of the Monte Carlo method. We use the RFPBE to study a two dimensional system as an example, in which the effect of ion fluctuation is clearly captured.
\end{abstract}

\maketitle

\section{Introduction}
The interaction between charged fluids and charged surfaces brings many novel phenomena, such as the condensation of DNA, aggregation of polymers, and like-charge attraction~\cite{1Hatlo,1Kanduc,1Lau1,1Lau2,1Lau3,1Linse,1Moreira,1Pelta1,1Pelta2,1Takahashi,1Tellez,1Yoshikawa}. It has been understood that the mechanism behind the phenomena is the fluctuation of ion density in the charged fluids~\cite{2Buyukdagli1, 2Podgornik1,2Podgornik2,2Naji,2Netz1,2Netz2,2Netz3,2Qiao,2Wang,Budkov2,2Wernersson1,2Wernersson2,2Buyukdagli2}. Even though the classical Poisson-Boltzmann equation (CPBE) has been widely used to study the ion distributions in charged fluids, it is well known that the CPBE is a mean field theory which averages the ion fluctuation and erases detail of the ion fluctuation ~\cite{3Attard,3Crocker,3Guldbrand,3Joensson,3Kjellander1,3Kjellander2,3Rouzina}. It is shown that such averaging loses the effect of the ion fluctuation and is not proper to describe the ion distributions in some cases, such as in the presence of ions with high valency and charge reversal of macromolecules~\cite{33Buyukdagli3}. In order to catch the effects of the ion fluctuation in the charged fluids, a theory beyond the CPBE needs to be developed. In this study, we derive a stochastic version of the Poisson-Boltzmann equation with a random field from the field theory. This Poisson-Boltzmann equation with a random field (RFPBE) recovers the ion fluctuation effects through the random field. Numerical solution to the RFPBE via Monte Carlo gives improved results about the ion fluctuation over CPBE. \\

As a powerful tool for studying the many-body systems, the field theory has been developed in the charged fluid to understand the ion fluctuations~\cite{2Buyukdagli1,2Podgornik1,2Podgornik2,2Naji,2Qiao,2Netz1,2Netz2,2Netz3,2Wang}. The field theory represents the partition function of a charged fluid in the form of functional integral. With the help of the Hubbard-Strotonovich transformation, an auxiliary field is introduced in the functional integral. All the physical properties of the fluid can be obtained from those of the auxiliary field. The saddle point solution to the functional integral is exactly the CPBE, in which the imaginary part of the auxiliary field is relevant to the electrostatic potential~\cite{2Naji,2Netz1,2Netz2,2Netz3,2Wang}. In order to catch the effects of the ion fluctuations, previous studies made expansion around the saddle point solution to the functional integral~\cite{2Netz1,2Netz2,2Netz3,2Wang,33Buyukdagli3,33Moreira,33Naji}. When the coupling between the charged fluid and the interface is extremely weak, a modified Poisson-Boltzmann equation has been obtained by the one-loop expansion of the functional integral~\cite{2Netz1,2Netz2,2Netz3}. Based on the modified Poisson-Boltzmann equation, a self-consistent theory has been developed with the help of the Gibbs variation~\cite{2Netz2}. It has been pointed out that the self-consistent theory is only valid for this weak coupling case~\cite{33Buyukdagli3}. It has also been pointed out that such Gibbs variation is on the first-order level for the perturbation variation~\cite{2Netz2}. When the coupling between the charged fluid and the interface is extremely strong, Viral expansion for the functional integral is a reasonable tool to  get the fluctuation effects~\cite{33Moreira}. Except for these two extremes, no general equation or even proper expansion has been developed yet from the functional integral when the coupling is in the intermediate regime~\cite{33Naji}.\\

The difficulty to get a general theory covering the whole range of the coupling lies in the fact that the ion fluctuation in the charged fluid itself have not been well described yet. Thus, we decouple the charged fluid from the charged interfaces and focus on the ion fluctuation in the charged fluid itself. The ion fluctuation represented by the real part of the auxiliary field can be transformed into a noise term by a secondary Hubbard-Strotonovich transformation. In this way, we derive the RFPBE from the field theory with the noise term added in the CPBE to capture the ion fluctuation. The coupling between the fluid and the interfaces is considered as boundary conditions in solving the RFPBE. Such treatment for the charged fluids is general and can be applied for various cases.\\

Recently, a stochastic differential equation was formulated to catch the fluctuation effect of the ion density in charged fluids~\cite{Poitevin1}. In that equation, the action of the functional integral is considered to be local and the conjugate field is integrated out at each point of space around the saddle point solution. In this way, the functional integral is transformed into one over positive density field only. A Langevin equation is then applied directly to get the stochastic equation, together with a reformulated Hamiltonian from the transformed functional integral. The stochastic equation in Ref.(\onlinecite{Poitevin1}) is different from the RFPBE in this paper. The former is for the charge density while the latter is for the electric field. We note that our theory is different from that of Ref.(\onlinecite{Poitevin1}) in two aspects. First, we do not utilize the locality of the field. Instead, we take the advantage of the real part of the complex auxiliary field. Second, our RFPBE is not obtained from the Langevin equation but from a secondary Hubbard-Strotonovich transformation. 
 
\section{Field theory}
The field theory has been well developed for charged fluids~\cite{2Naji,2Netz1,2Netz2,2Netz3,2Wang}. To be self-contained, we present the field theory in this section.\\

We consider a charged fluid confined by solid boundaries. To illustrate the field theory clearly, the fluid consists of only two ion species with opposite charges. For the positive charges, the charge value of each ion is denoted by $z_+$ and the ion number is by $N_+$. For the negative charges, they are $z_-$ and $N_-$ respectively. We also denote the elementary positive charge by $e$. Generally, the total net charge $ez_+N_+-ez_-N_-$ in the fluid could be nonzero and can be compensated by external charges distributed in the solid boundaries. Thus, the total net charge of the whole system including the fluid and boundaries still could be neutral. We ignore the structures of the ions and the dielectric difference between the fluid and the solid boundaries. Therefore, the steric effect and image-charge effect are not considered in our theory. The dielectric constant of the system is denoted by $\epsilon$.

\subsection{Partition function} 
We start from the canonical partition function of the system
\begin{align}
Q=\frac{1}{N_+!N_-!\lambda_+^{N_+}\lambda_-^{N_-}}\int \prod_{i=1}^{N_+}d\vec{r}_i\prod_{j=1}^{N_-}d\vec{r}_je^{-\beta H+\int d\vec{r} h(\vec{r})\rho(\vec{r})}.
\end{align}
Here, $\lambda_+$ and $\lambda_-$ are the de Broglie wavelengths for the positive charges and the negative ones respectively. $\vec{r}_i$ is the position vector of the i-th ion and $d\vec{r}_i$ is the infinitesimal volume for the integration. $\beta$ is the inverse temperature. In the exponent, the Coulomb energy is
\begin{align}
H=\frac{e^2}{2}\int d\vec{r} d\vec{r}' \rho(\vec{r}) C(\vec{r},\vec{r}')\rho(\vec{r}').
\end{align}
It is expressed in terms of the Green function $C(\vec{r},\vec{r}')$, which satisfies the equation $-\nabla_r \cdot [\epsilon \nabla_r C(r,r')]=\delta(r-r')$. The ion density of the positive charges is denoted by $c_+$ and that of the negative ones by $c_-$, with $c_{\pm}(\vec{r})=\sum_{i=1}^{N_{\pm}}\delta(\vec{r}-\vec{r}_i)$. The net charge density then can be expressed by $e\rho(\vec{r})=e\sigma(\vec{r})+e[  z_+c_+(\vec{r})- z_-c_-(\vec{r})]$ with $e \sigma(\vec{r})$ the external charge density in the solid boundaries. $\rho(\vec{r})$ is the net ion density with the elementary positive charge for each ion. The function $h(\vec{r})$ introduced in the exponent is to generate the averaged ion density through  $\langle \rho(\vec{r}) \rangle= \partial \ln Q/\partial h(\vec{r}) |_{h=0}$. \\

Now we apply the Hubbard-Stratonovich transformation on the partition function $Q$ and introduce an auxiliary field $\phi$. Denoting the imaginary unit by $i$, we obtain 
\begin{align}
& Q=\frac{1}{Z_c}\frac{1}{N_+!N_-!\lambda_+^{N_+}\lambda_-^{N_-}}\int [D \phi]\cdot \nonumber \\
& e^{- \int d\vec{r} \left\{\frac{1}{2}[\nabla_{\vec{r}} \phi(\vec{r})]\epsilon[\nabla_{\vec{r}}\phi(\vec{r})]+[i e\sqrt{\beta} \phi(\vec{r})-h]\sigma(\vec{r})\right\}}\Lambda_+^{N_+}\Lambda_-^{N_-}
\end{align}
with
\begin{align}
&\Lambda_+=\int d\vec{r} e^{-z_+[i e\sqrt{\beta} \phi(\vec{r})-h(\vec{r})]},\nonumber \\
&\Lambda_-=\int d\vec{r} e^{z_-[i e\sqrt{\beta} \phi(\vec{r})-h(\vec{r})]},\nonumber \\
&Z_c=\int [D \phi] e^{- \left\{\frac{1}{2}\int d\vec{r} d\vec{r}' \phi(\vec{r}')C^{-1} \phi(\vec{r})\right\}}.
\end{align}
We introduce fugacities $\mu_+$ for positive charges and $\mu_-$ for negative ones. The grand canonical partition function then is obtained through $\Xi=\sum _{N_+=0}^{\infty}\sum _{N_-=0}^{\infty}Q(N_+,N_-)\mu_+^{N_+} \mu_-^{N_-}$. For convenience, we use the Bjerrum length $l_B'=e^2\beta/\left(4\pi \epsilon \right)$ in standard units and introduce our length scale $l_B=4\pi l_B'$ to scale all the lengths. The Bjerrum length $l_B'$ equals $6.96\AA$ for water at room temperature. We rewrite $\vec{r}/l_B$ by $r$, $e\sqrt{\beta}\phi$ by $\phi$ and $\sigma l_B^3$ by $\sigma$. We also define $w_+=\mu_+ l_B^3/\lambda_+$, $w_-=\mu_- l_B^3/\lambda_-$. Then the grand canonical partition function $\Xi$ reads
\begin{align}
\label{Xi}
\Xi=\frac{1}{Z_c}\int [D \phi]e^{\int dr \left[ A_1(r)+A_2(r) + A_3(r)+A_4(r) \right]}.
\end{align}
with
\begin{align}
\label{A}
&A_1(r)=-\frac{1}{2}[\nabla_r \phi(r)]^2, ~~~~~~~~A_2=[h(r)-i\phi(r)]\sigma,\nonumber \\
&A_3(r)=w_+e^{-z_+[i\phi(r)-h(r)]},~~ A_4(r)=w_-e^{z_-[i \phi(r)-h(r)]}.
\end{align}
Clearly, the partition function $\Xi$ is a functional of the auxiliary field $\phi(r)$.

\subsection{Auxiliary field $\phi(r)$}
The auxiliary field $\phi(r)$ can be considered as field fluctuations around classical field~\cite{2Netz1}. The classical field lays on the imaginary axis in complex contour while the field fluctuations are real and orthogonal to the imaginary axis. Thus, the auxiliary field $\phi(r)$ can be written in the form of $\phi=\phi_R+i \phi_I$ with $\phi_I$  the classical field solution and $\phi_R$ referred to the field fluctuations. In the one-loop method suggested in Ref.(\onlinecite{2Netz1}), the classical field is the saddle point solution to the partition function $\Xi$ and the final solution to the electrostatic field is obtained by modifying the classical field with the field fluctuations expanded around the saddle point solution. In the one-loop method, the saddle point solution is exactly the mean field solution to the CPBE. Similar to Ref.(\onlinecite{2Netz1}), we will expand the field fluctuation $\phi_R$ around the classical field $\phi_I$. And then we transform the field fluctuation into a multiplicative noise added in the CPBE to get the  RFPBE. The difference of our treatment to the one-loop expansion is that the classical field $\phi_I$ used in our theory is not the saddle point solution to the partition function $\Xi$. Instead, it has been modified by the field fluctuation, and need to be solved from the RFPBE.\\

We specify more for the physical meaning of the auxiliary field. The term relevant to $\phi_R$ in $A_1$ of Eq.(\ref{A}) is quadratic $-\frac{1}{2}[\nabla_r \phi_R(r)]^2$. In the functional integral Eq.(\ref{Xi}), the quadratic term of $\phi_R$ appears in the exponent of    
\begin{align}
\label{B}
B(\phi_R)= e^{-\frac{1}{2}\int dr [\nabla_r \phi_R(r)]^2}.
\end{align}
To show the physical meaning of $B(\phi_R)$, we take one dimension fluid as an example. We discretize $B(\phi_R)$ as
\begin{align}
\label{B1}
B(\phi_R)= \lim \limits _{\Delta \rightarrow 0}e^{-\sum_n \frac{(\phi_{R,n+1}-\phi_{R,n})^2}{2\Delta}}
\end{align}
with $\Delta$ the width of discrete mesh. $\phi_{R,n}$ means $\phi_R(r)$ at the $n$-th lattice and behaves as a Gaussian noise. In averaging, $\phi_R$ takes zero at the saddle point solution to the partition function $\Xi$.\\

We set $h=0$ and solve the functional derivative equation $\delta \Xi/\delta \phi=0$ to get the saddle point solution that reads ( neglecting the $r$ variable in $\sigma$ and $\phi$ fields)
\begin{align}
\label{clasPB}
-\nabla ^2[i\phi]=\sigma+z_+w_+ e^{-z_+[i\phi]}-z_-w_- e^{z_-[i\phi]}.
\end{align}
In order to understand the above equation, we apply $\langle \rho(\vec{r}) \rangle= \partial \ln \Xi / \partial h(\vec{r})|_{h=0}$  to get the ion density of the charged fluid
\begin{align}
\label{density}
\langle \rho(\vec{r}) \rangle = \langle \sigma+z_+w_+ e^{-z_+[i\phi]}-z_-w_- e^{z_-[i\phi]} \rangle.
\end{align}
In the above equation, $\sigma$  is the density of ions in solid boundaries. The second term means the density of the positive-charged ions in the fluid with unit charge per ion, and the third term is the density of negative-charged ions. Thus, the right hand side of Eq.(\ref{clasPB}) is understood as the net ion density for one measure of the ion fluctuation. If we drop off $\phi_R$ because of $\phi_R=0$ at the saddle point solution and only keep $i\phi_I$, the CPBE is recovered from Eq.(\ref{clasPB}). The classical filed $-\phi_I$ is then interpreted as the electrostatic potential.\\

\section{RFPBE}
We derive the RFPBE for two goals. The first goal is to get a general equation covering the whole range from the weak coupling to the strong coupling to catch the effect of the ion fluctuation. It is known that the one-loop theory and self-consistent theory are valid for the weak coupling case~\cite{33Buyukdagli3}. In order to achieve this goal, we decouple the fluid from the charged interfaces and focus on the ion fluctuation in the charged fluid itself. The coupling between the charged interfaces and the fluid is set as the boundary conditions to solve the RFPBE. Then, the coupling parameter well defined in the one-loop theory and the self-consistent theory is absent in our RFPBE and absorbed into the boundary conditions. In this way, the  RFPBE is general and flexible for various cases to catch the effect of ion fluctuations in the fluid. The second goal is to get a proper treatment to catch ion fluctuation in high dimensional systems by using the stochastic nature of the  RFPBE. For the well known self-consistent theory, various approximations have to be made to simplify the numerical calculations ~\cite{1Hatlo,2Buyukdagli1}. Comparably, by taking the advantage of the Monte Carlo method, no approximation is needed in solving the RFPBE, which makes the RFPBE feasible for high dimensional systems. We will study a two-dimensional system as an example later. Now we derive the RFPBE.

\subsection{Equation}
We set $h=0$ and substitute $\phi=\phi_R+i\phi_I$ into $A_3$ and $A_4$ of Eq.(\ref{A}). Thus, we have $A_3(r)=w_+e^{-z_+[i\phi_R-\phi_I]}$ and $A_4(r)=w_-e^{z_-[i\phi_R-\phi_I]}$. Due to the Gaussian distribution of $\phi_R$ in Eq.(\ref{B1}), series expansion can be applied on $A_3$ and $A_4$ in terms of $\phi_R$. The higher order terms in the expansion bring the more accurate result in the calculation. Here, we only want to achieve the two goals mentioned in the beginning of this section. The trade-off is to decrease the accuracy in the calculation. Thus, we expand $A_3$ and $A_4$ only to the second order, reading
\begin{align}
\label{A3}
&A_3(r)=w_+e^{z_+\phi_I}\{1-z_+i\phi_R+\frac{z_+^2}{2}[i\phi_R]^2\},\nonumber \\
&A_4(r)=w_-e^{-z_-\phi_I}\{1+z_-i\phi_R+\frac{z_-^2}{2}[i\phi_R]^2\}.
\end{align}
We substitute Eq.(\ref{A3}) in Eq.(\ref{Xi}), and rewrite $\Xi$ as
\begin{align}
\label{Xi1}
\Xi=\frac{1}{Z_c}\int [D \phi]e^{\int dr \{T_1+i\phi_R T_2+\frac{1}{2}[i\phi_R]^2T_3\}}
\end{align}
with
\begin{align}
\label{T}
&T_1=\frac{1}{2}[\nabla \phi_I]^2-\frac{1}{2}[\nabla \phi_R]^2+\sigma \phi_I+w_+ e^{z_+\phi_I}+w_-e^{-z_-\phi_I}, \nonumber\\
&T_2=\nabla^2\phi_I-\sigma-z_+w_+e^{z_+\phi_I}+z_-w_-e^{-z_-\phi_I},\nonumber \\
&T_3=z_+^2w_+e^{z_+\phi_I}+z_-^2w_-e^{-z_-\phi_I}.
\end{align}
Here, integral by parts has been applied to transform $\nabla \phi_R\cdot \nabla \phi_I$ to  $- \phi_R\nabla^2 \phi_I$. Since $T_3$ is positive, it is $[i\phi_R]^2=-\phi_R^2$ on the exponent of Eq.(\ref{Xi1}) that guarantees the low probability of the ion fluctuation with large $\phi_R$, which is consistent to the Gaussian distribution of $\phi_R$ in Eq.(\ref{B1}) and our treatment that $\phi_R$ is small for the expansion of Eq.(\ref{A3}). We note that if the higher order terms are used in the expansion of Eq.(\ref{Xi1}), the random field obtained in the RFPBE then is not Gaussian. In such a case, more precise tools are required to treat the random field. However, the higher order terms may bring some effects to the fluctuation, but will not dominate the fluctuations because of their small magnitudes. So the expansion to the second order, or the Gaussian noise in Eq.(\ref{Xi1}), is proper to catch the main contribution in the ion fluctuations. Since ion distribution in the fluid mainly depends on the mean field potential $\phi_I$, and the fluctuation $\phi_R$ only perturbs the distribution, we have $|\nabla \phi_R|\ll|\nabla \phi_I|$ and drop off the second term in $T_1$ of Eq.(\ref{T}). In this way, only $i\phi_R T_2+\frac{1}{2}[i\phi_R]^2T_3$ in the exponent of Eq.(\ref{Xi1}) is relevant to $\phi_R$. It could be understood that if the condition $|\nabla \phi _R|\ll|\nabla \phi _I|$ is violated, say $|\nabla \phi _R|>\nabla \phi _I|$, then the fluctuation force $|\nabla \phi _R|$ applied on the ions will pull the ions away from their mean field positions and scatter the ions randomly in the fluid. Under this condition, no mean field solution can be obtained. Such situation could happen for non-equilibrium states such as the fluid is in convection, but not for the equilibrium systems that we study. Therefore, the condition  $|\nabla \phi _R|<<|\nabla \phi _I|$ is applicable. \\

Before we derive the RFPBE, we first study the property of the term  $i\phi_R T_2$ by neglecting the quadratic term $\frac{1}{2}[i\phi_R]^2T_3$ in the exponent. The functional integral over $\phi_R$ in $\int [D\phi]e^{\int d r i\phi_R T_2}$ of Eq.(\ref{Xi1}) leads to a functional $\delta(T_2)$. That means $\phi_I$ is governed by a differential equation $T_2=0$, which is exactly the mean field theory CPBE. In order to consider the effects of the ion fluctuation, we should keep the quadratic term $\frac{1}{2}[i\phi_R]^2T_3$ in the exponent and figure out the stochasticity from this term.\\

We discretize the quadratic term in space by
\begin{align}
e^{\int dr \frac{T_3}{2}[i\phi_R]^2}=\lim \limits _{\Delta \rightarrow 0}e^{\sum_n \frac{\Delta T_{3,n}}{2}[i\phi_{R,n}]^2}.
\end{align}
Here, $\Delta$ is the volume of discrete mesh in 3D system, which can be reduced to the area in 2D and the width in 1D. $T_{3,n}$ means the value of $T_3$ at the $n$-th lattice site, similar to the notation of $\phi_{R,n}$. We introduce a variable $\alpha$ at each lattice site and apply the Hubbard-Stratonovich transformation again to get
\begin{align}
e^{\sum_n \frac{\Delta T_{3,n}}{2}[i\phi_{R,n}]^2}=\int [D\alpha] e^{-\sum_n\frac{\alpha_n^2}{2}}e^{\sum_ni\sqrt{\Delta T_{3,n}}\phi_{R,n}\alpha_n}
\end{align}
with $[D\alpha]=\prod_n  \frac{d \alpha_n}{\sqrt{2\pi}}$. The partition function Eq.(\ref{Xi1}) now becomes
\begin{align}
\label{GXI}
\Xi=&\frac{1}{Z_c}\int [D \phi][D\alpha]e^{\sum_n[-\frac{\alpha_n^2}{2}+\Delta T_{1,n}]}\nonumber\\
& \times e^{\sum_n i\phi_R,n[\sqrt{\Delta T_{3,n}}\alpha_n+\Delta T_{2,n}]}.
\end{align}
The functional integral over $\phi_R$ in $\int [D\phi] e^{\sum_n i\phi_R,n[\sqrt{\Delta T_{3,n}}\alpha_n+\Delta T_{2,n}]}$ of the above $\Xi$ leads to a functional $\delta (\sqrt{T_{3,n}}\frac{\alpha_n}{\sqrt{\Delta}}+ T_{2,n})$ at each lattice, just like what we have done on the term  $i\phi_R T_2$ in the last paragraph. The $\delta$ functional means the system is governed by a equation $ \sqrt{T_{3,n}}\frac{\alpha_n}{\sqrt{\Delta}}+ T_{2,n}=0$ at each lattice. We map the temporal noise in the theory of stochastic process to a spatial noise in our study by introducing a variable $\eta_n$ to replace $\alpha_n/\sqrt{\Delta}$ in the $\delta$ functional. Further, we use $\psi$ to replace $-\phi_I$ in the equation to represent the electrostatic potential. Finally, we obtain the  RFPBE explicitly as
\begin{align}
\label{SPB}
&-\nabla^2\psi=\sigma+z_+w_+e^{-z_+\psi}-z_-w_-e^{z_-\psi}\nonumber \\
&-\sqrt{z_+^2w_+e^{-z_+\psi}+z_-^2w_-e^{z_-\psi}}~\eta.
\end{align}
Here, $\eta$ is a Gaussian noise due to the Gaussian distribution $e^{-\frac{\alpha^2}{2}}$ of $\alpha$ in Eq.(\ref{GXI}). It is a function of position in the charged fluid with correlation $\langle \eta(r)\eta(r') \rangle=\delta(r-r')$. Except the last term, the other part in Eq.(\ref{SPB}) is exactly the CPBE.  The last noise term represents the density fluctuation of ions. There are two sources contributing to the  noise term. The first source is the reservoir. Since the functional integral in the field theory is transformed from the grand canonical partition function, the computational domain actually is attached to a reservoir of the charged fluid. The ions can flow from the reservoir to the computational domain, or from the domain to the reservoir inversely, which induces the fluctuation of the ion density in the domain. In equilibrium, the fluctuation is averaged to be zero but locally it is nonzero. The second source to the noise term is the Brownian motion of the ions. In the liquid, the ions are mobile and can move from place to place due to the interactions by dielectric molecules, such as collision or Lennard-Jones interaction. Such Brownian motion brings the density fluctuation of ions in the domain. According to the many-body theory, the ions in systems with long-range interaction must be screened. Such screening effect has been reflected by the screening function multiplying the noise. For the general case where the charged fluid contains various ionic species ($k=1,2,...$), the RFPBE reads
\begin{align}
\label{SPB1}
-\nabla^2\psi = \sigma+\sum_k a_k z_k w_k e^{-a_k z_k \psi}-\sqrt{\sum_k z_k^2 w_k e^{-a_k z_k \psi}}~\eta
\end{align}  
with $a_k=+1$ for positive charges and $a_k=-1$ for  negative charges. For fluids containing various types of ions, correlations between the different types of ions are reflected by the multiplicative noise term in Eq.(\ref{SPB1}).\\

The Gaussian noise in the last term of Eq.(\ref{SPB1}) originates from the field fluctuation $\phi_R$ and is the source to the motion of ions. The screening function multiplying the Gaussian noise comprises a multiplicative noise, which captures the effect of the ion fluctuation in the charged fluid. Any motion of ions due to the ion fluctuation changes the electrostatic field and the changing of the field feeds back to the distribution of ion density through the multiplicative noise, which brings into a new electrostatic field until the system reaches a equilibrium state.

\subsection{Charge conservation}
In Eq.(\ref{SPB1}), the parameters $w_k$ have not been determined yet. According to the CPBE, $w_k$ should be the bulk ion density of the $k$-th ionic species. However, such treatment brings a problem that charges in the fluid are not conserved. The total ion number $\int drw_ke^{-a_kz_k\psi}$ is not always equal to the total ion number $\int dr w_k$ in bulk. This charge conservation problem has been pointed out by one of the authors and others in a recent work~\cite{4Wan}. Several methods have been proposed to guarantee the charge conservation~\cite{4Wan}. Here, we adopt the following method for the numerical calculation~\cite{4Wan, 4Lee}.\\

The charge conservation in the fluid means the total ion number must not be changed no matter what is the distribution of the ion density. We note the bulk density by $M_k$ for the $k$-th ionic species. The charge conservation requires $\int dr w_ke^{-a_kz_k\psi}=M_k \int dr$. Since $w_k$ depends on the chemical potential that is constant in equilibrium, we have $w_k$ as
\begin{align}
\label{wn}
w_k=\frac{M_k \int dr}{\int dr e^{-a_kz_k\psi}}.
\end{align}
The integrations are over the total computational domain.

\subsection{Boundary condition}
The coupling between the interface and the fluid is taken into account by the boundary condition(BC) of the RFPBE. Three BCs have been used~\cite{4Wan}. The first BC is the Dirichlet BC, in which the electrostatic potentials are fixed at the interfaces between the fluid and the solid boundaries. Such BC is also known as $\zeta$ potential BC in colloidal science. The second BC is the Neumann BC, in which the derivative of electrostatic potential  with respect to spatial coordinate is fixed at the interfaces. The last BC is the Robin BC, which is the mixture of the Dirichlet BC and Neumann BC.\\

It has been observed in some experiments that the BCs at the interfaces are not fixed but depend on the experimental conditions in bulk. For example, $\zeta$ potential or the interface charges vary when the ion densities or PH values in the fluid are changed. To treat such variation of BCs, several models have been proposed, such as the charge regulation model and the potential trap model ~\cite{4Wan, 5Behrens,5Camara,5Carnie,5Gentil}.\\

In this study, for illustration purpose, we only apply the Dirichlet BC to our  RFPBE. We propose a Monte Carlo method based on the path integral representation to solve the RFPBE with the Dirichlet BC. The studies on other BCs are not in the scope of this work. 

\section{Numerical calculation}
The RFPBE has a multiplicative noise in Eq.(\ref{SPB1}). The white noise $\eta$ is coupled to a function of the electrostatic potential, which means any change of the electrostatic potential can be fed back to itself through the noise. To solve this equation, we propose a Path Integral Monte Carlo (PIMC) method. It is a Monte Carlo method based on path integral representation~\cite{MCBinder}. For clarity, we first reduce our system to one dimension along $x$ axis. The numerical calculation is then carried out for a two dimensional system. The PIMC method described in this study can also be easily extended to the three dimensional system.

\subsection{Path Integral representation}
 We propose the PIMC for a general stochastic equation
\begin{align}
\label{spbgen}
-\frac{d^2 \psi}{dx^2}=f(\psi)-g(\psi)\eta.
\end{align}
We introduce a function $K(x)=\frac{1}{g}\frac{d \psi}{dx}$ to transform the above equation to two first order differential equations
\begin{align}
\label{equs}
&\frac{dK}{dx}+K^2 \frac{dg}{d\psi}+\frac{f(\psi)}{g(\psi)}-\eta=0, \nonumber \\
&\frac{d \psi}{dx}=gK.
\end{align}
After such transformation, the noise term appearing in the stochastic equation is additive instead of multiplicative. We use $h(K, \psi)$ to represent $K^2 \frac{dg}{d\psi}+\frac{f(\psi)}{g(\psi)}$  for simple notation and note $\eta dx$ by $dW$. Then, we discretize the first equation of Eq.(\ref{equs}) in Stratonovich sense as
\begin{align}
K_n-K_{n-1}+\frac{(h_n+h_{n-1})\Delta}{2}=W_n-W_{n-1}.
\end{align}
The Jacobian determinant for the variable transformation reads
\begin{align}
\frac{dW_n}{d K_n}=1+K_n \frac{d g_n}{d \psi_n}\Delta \approx  e^{K_n \frac{d g_n}{d\psi_n}\Delta}.
\end{align} 
Here, $\frac{d g_n}{d \psi_n}$ means the derivative of the function $g(\psi)$ with respect to $\psi$ at the $n^{th}$ lattice. The probability for one electrostatic potential path starting from $\psi_0$ at $x_0$ to $\psi_N$ at $x_N$ is
\begin{align}
\label{PI}
&P(\psi_N, x_N|\psi_0, x_0)\nonumber \\
&=\int [D\psi][DW]\delta(\psi-\psi_0)\delta(\psi-\psi_N )\delta \left( \frac{d \psi}{dx}-gK \right) \nonumber \\
&~~~~~~~\cdot e^{-\int _{x_0}^{x_N}\frac{1}{2}\eta^2dx}\nonumber \\
&=\int [D\psi][DK]\delta(\psi-\psi_0)\delta(\psi-\psi_N)\delta\left( \frac{d \psi}{dx}-gK  \right)  \nonumber \\
&~~~~~~~\cdot e^{\int dx \left\{ K \frac{d g}{d\psi}-\frac{1}{2}[\frac{dK}{dx}+K^2 \frac{dg}{d\psi}+\frac{f(\psi)}{g(\psi)}]^2 \right\} }.
\end{align}
Here, $[DW]=\prod _n \frac{dW_n}{\sqrt{2\pi \Delta}}$, $[D \psi]=\prod _n d \psi_n$ and $[DK]=\prod _n \frac{dK_n}{\sqrt{2\pi \Delta}}$. The action in the path integral is represented by $L$ as
\begin{align}
\label{action}
L=\int dx \left\{ K \frac{d g}{d\psi}-\frac{1}{2} \left[\frac{dK}{dx}+K^2 \frac{dg}{d\psi}+\frac{f(\psi)}{g(\psi)} \right]^2 \right\},
\end{align}
which will be used for the Metropolis algorithm later. We note that Eq.(\ref{PI}) has an intrinsic relation to Eq.(\ref{GXI}) by considering each component of the partition function Eq.(\ref{GXI}) as the probability of path. But the direct relation between the two path probabilities is still lack.

\subsection{PIMC}
For our reduced one-dimensional fluid, the two interfaces are set at $x=0$ and $x=b$ respectively. The charged fluid is confined in $0<x<b$. For illustration, the charged fluid contains only one ionic species and the ions are positive. For the fluid, we have $\sigma=0$ in Eq.(\ref{SPB1}). Considering the charge conservation Eq.(\ref{wn}), we express the  RFPBE in an explicit form according to Eq.(\ref{SPB1}). It reads 
\begin{align}
\label{PIMCspb}
-\frac{d^2 \psi}{dx^2}=\frac{zbMe^{-z\psi}}{\int dx e^{-z\psi}}-\vartheta \sqrt{ \frac{z^2bMe^{-z\psi}}{\int dx e^{-z\psi}}}~\eta.
\end{align}
In this equation, the charge is always conserved for the solution. Comparing to Eq.(\ref{spbgen}), we have $f(\psi)=\frac{zbMe^{-z\psi}}{\int dx e^{-z\psi}}$ and $g(\psi)=\vartheta \sqrt{ \frac{z^2bMe^{-z\psi}}{\int dx e^{-z\psi}}}$. A parameter $\vartheta$ is introduced in $g(\psi)$ to control the intensity of noise. $\vartheta=0$ is for the CPBE while $\vartheta=1$ is for the  RFPBE. Practically, we will take $\vartheta=0.01$ instead of $\vartheta=0$ for the CPBE in the numerical study.  \\

We discretize the computational domain from $x=0$ to $x=b$ by $N$ slices. The discrete lattice sites are indexed from $0$ to $N$. The $0$-th and the $N$-th lattice sites are fixed at the boundaries and have their potentials of $\psi_0$ and $\psi_N$, respectively. Before sampling the paths, we need to initialize the potential path for the start. It does not matter which potential path is used for the initialization. The Markovian process of noise will lose its memory in the sampling.\\

We denote the old potential path by $\psi^{old}$ and the new one by $\psi^{new}$. In the sampling, for site $n$ we generate $\psi^{new}$ on this site by $\psi_n^{new}=\psi_n^{old}+\delta \psi (2\omega -1)$. Here, $\delta \psi$ is the step magnitude for the moving of $\psi$ and $\omega$ is a random number with uniform distribution in the range $[0,1]$. Now the new path $\psi^{new}$ is generated from $\psi^{old}$ by replacing $\psi_n^{old}$ with $\psi_n^{new}$ and keeping $\psi_{i\neq n}^{old}$ intact. Similarly, we can sample numerous $K$ paths for the calculation. However, the accepting ratio of the new $K$ paths is governed by the factor $\delta(\frac{d \psi}{dx}-gK)$ in the conditional probability Eq.(\ref{PI}). That means the new $K$ paths can be accepted only if they satisfy  $K=\frac{1}{g}\frac{d \psi}{d x}$. Otherwise, they will be rejected. Therefore, we can get the new $K$ path directly from $\psi^{new}$  through the second equation in Eq.(\ref{equs}).\\

After obtaining the new path $\psi^{new}$, we make the integration of $\int dx e^{-z\psi^{new}}$ over the system as the denominator in Eq.(\ref{PIMCspb}). In this way, the functions $f(\psi^{new})$ and $g(\psi^{new})$ in Eq.(\ref{spbgen}) are updated. This step is very important for the charge conservation respecting each new path. By using the new functions $\psi$, $K$, $f$ and $g$, we calculate the action $L^{new}$ according to Eq.(\ref{action}).\\

Since the random numbers generated for $\psi$ are uniform, the detailed balance requires that the accepting ratio for the new path is
\begin{align}
A_r=\frac{e^{L^{new}}}{e^{L^{old}}}.
\end{align} 
Then Metropolis algorithm is applied for the accepting ratio~\cite{MCBinder}. Explicitly, the newly generated path $\psi^{new}$ is accepted with probability $\min \{1,A_r\}$. That is, if $A_r \ge 1$, the new path $\psi^{new}$ is accepted. If $A_r < 1$, we generate a uniform random number $A_q$ in the range $[0,1]$. If $A_r > A_q$ is true, we accept the new path. If not, we reject the new path and copy the old path for the update.\\

In practice, the PIMC runs from the $1^{st}$ lattice to the $(N-1)^{th}$ lattice. And then return to the  $1^{st}$ lattice again for many cycles. After reaching the equilibrium, the paths then are used for statistical calculations.

\section{Results and discussions}
In this section, we will compare the results from CPBE and  RFPBE. In our RFPBE, the electrostatic potential has been renormalized by $1/\beta$ which is $26 meV$ at room temperature. All the lengths have been scaled by $l_B$, which is $4\pi$ times the Bjerrum length $l_B'$ and equals $87.46\AA$ for water at room temperature. As mentioned above,  Eq.(\ref{PIMCspb}) is used in this study with $\vartheta=0.01$ for the CPBE and $\vartheta=1$ for the  RFPBE.\\

We take $M=0.04$ and $z=1$, which corresponds to $1\times 10^{-4} mol/L$ of electrolyte and the Debye length of such system is $5 l_B$. The length $b$ of the charged fluid is set to be $20$ and the potentials at boundaries are set to be $\psi_0=-2.0$ and $\psi_N=0.0$ respectively. We present the electrostatic potentials in Fig.1(a). For the CPBE, the potential goes up from $\psi_0=-2.0$ at $x=0$ to $\psi_N=0.0$ at $x=b$ as a continuous convex function. However, the RFPBE shows that the potential goes up from $\psi_0=-2.0$ as a convex function and then behaves as a concave function before reaching $\psi_N=0.0$. There exists a segment in the RFPBE curve at which the convex function transits to a concave function. The existence of the transition segment is the feature of the nonlinear effects of the ion fluctuation captured by the RFPBE and it distinguishes the RFPBE from the CPBE. The transition segment is in the domain instead of close to the interfaces and the potential gradient is small in the segment compared to the large potential gradient close to the interfaces. At the center of the domain, the deviation between the two potentials is calculated to be $13.52 meV$ in the real unit.\\

In Fig.1(b), we present the ion density $\langle \rho \rangle = \langle \frac{zbMe^{-z\psi}}{\int dx e^{-z\psi}} \rangle$ . 
\begin{figure}[!hbp]
\centering
\includegraphics[width=0.5\textwidth]{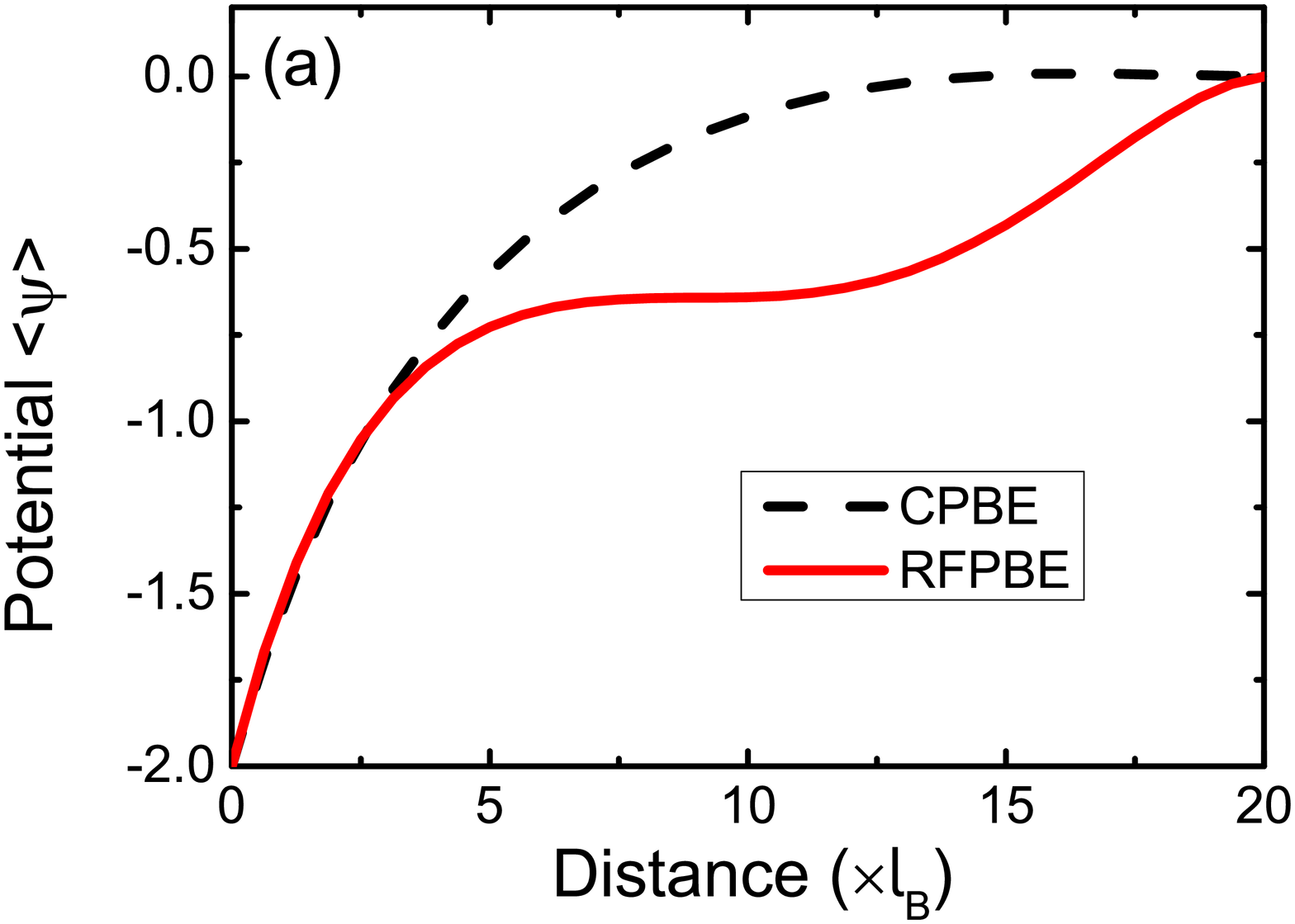}
\includegraphics[width=0.5\textwidth]{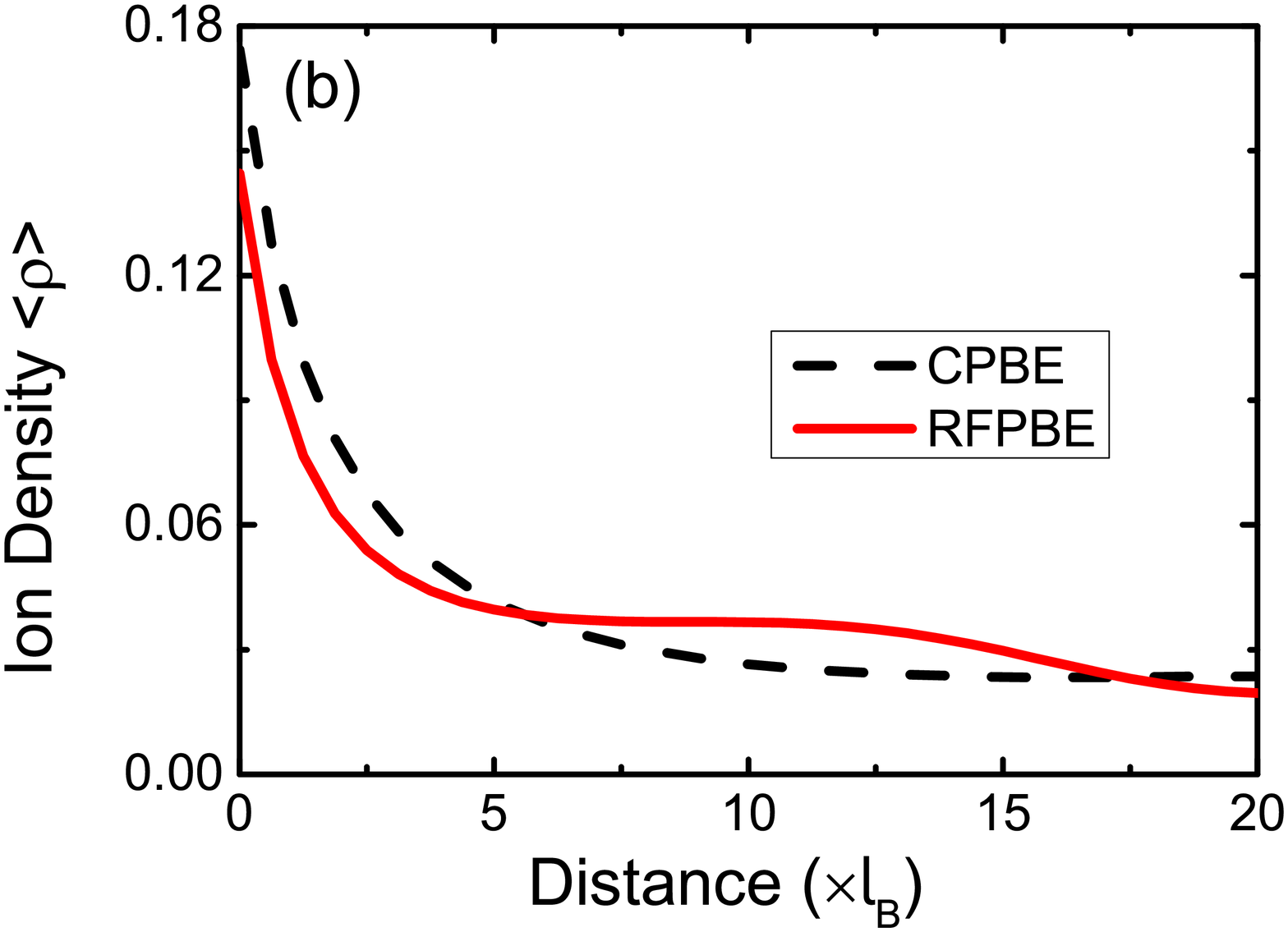}
\caption{Comparison of the CPBE and the RFPBE results. The solid lines are for the RFPBE results while the dashed lines for the CPBE results. (a )Electrostatic potentials. (b) Ion densities. }
\end{figure}
It shows that the positive charges of both the CPBE and the RFPBE concentrate close to the interface at $x=0$ due to the negative potential $\psi_0$ and then decay with $x$ up to the interface at $x=b$. But we can find detailed differences between the CPBE and RFPBE. In the CPBE result, the ion density decays in the same manner as a continuous concave function. However, in the RFPBE curve the ion density experiences a transition from a concave function to a convex function. Compared to the CPBE result, the RFPBE result has a lower ion density at the interface $x=0$ but a higher density in the center of domain. The two curves cross at the Debye length $5l_B$. In Fig.1(b), the deviation of the ion density at the center of the domain is calculated to be $1.01\times 10^{-6} mol/L$ in the real unit.\\

The ion distribution is determined by the competition between the ion fluctuation and the electrostatic force applied by the interfaces. The ion fluctuation increases the entropy of the system and tends to disperse the ions uniformly in the whole domain. Eventually, the dispersion of the ions into the domain is balanced by the electrostatic force provided by the boundaries, forming the final ion distribution. In the CPBE, the ion fluctuation is averaged as the mean field effect balanced by the electrostatic force. Such averaging misses the nonlinear effect of the ion fluctuation. The multiplicative noise in the  RFPBE Eq.(\ref{SPB1}) has captured the nonlinear effect of the ion fluctuation around the mean field solution. Such nonlinear effect strengthens the dispersion of ions into the domain, leading to the transition segment in the RFPBE curve of Fig.1(b).\\

Since the nonlinear effect of the noise plays so important role in the ion distribution, we control the noise intensity by varying the control parameter $\vartheta$ and present the results in Fig.2 . The larger $\vartheta$, the stronger intensity of the multiplicative noise is.
\begin{figure}[!hbp]
\centering
\includegraphics[width=0.4\textwidth]{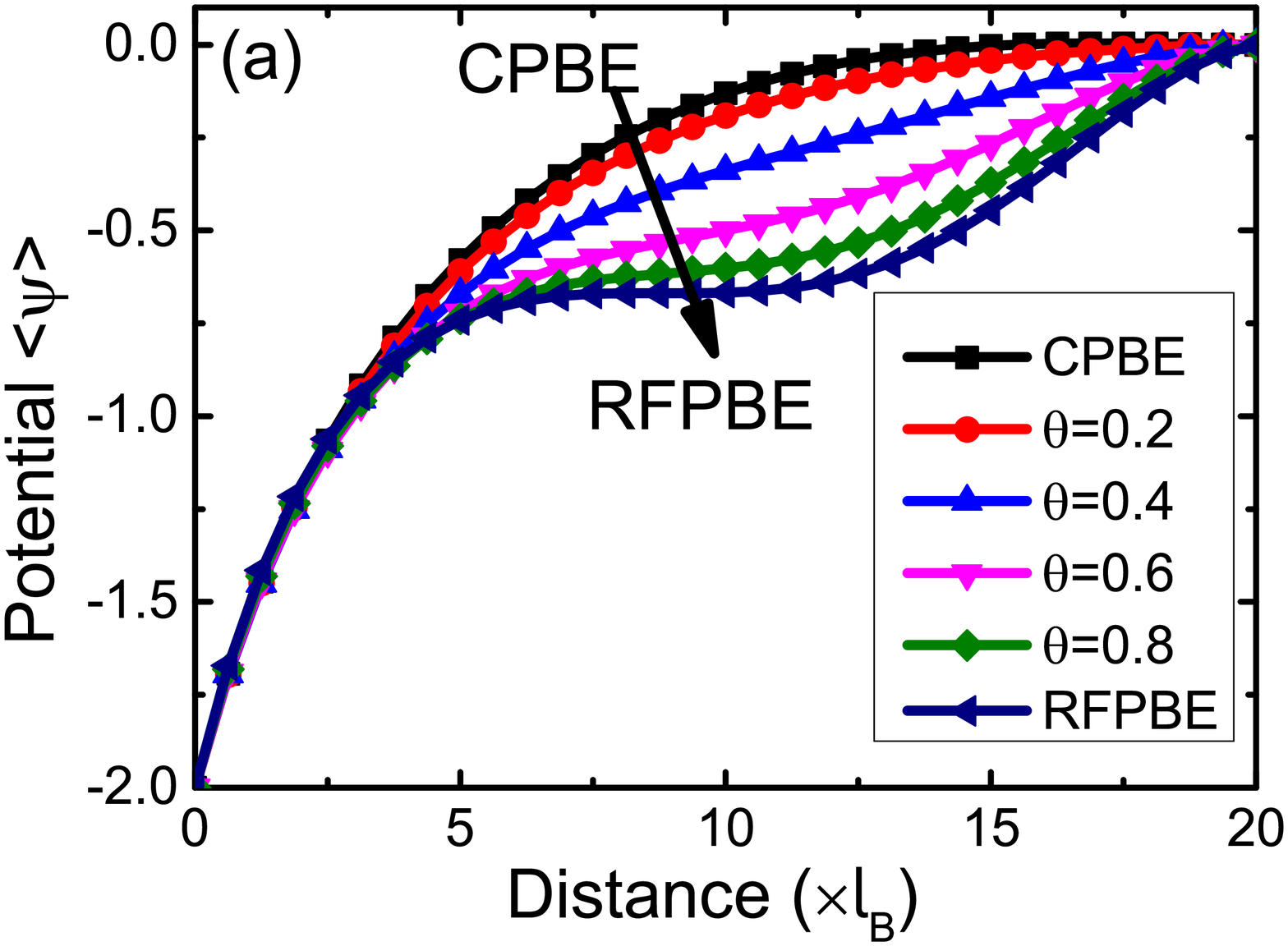}
\includegraphics[width=0.4\textwidth]{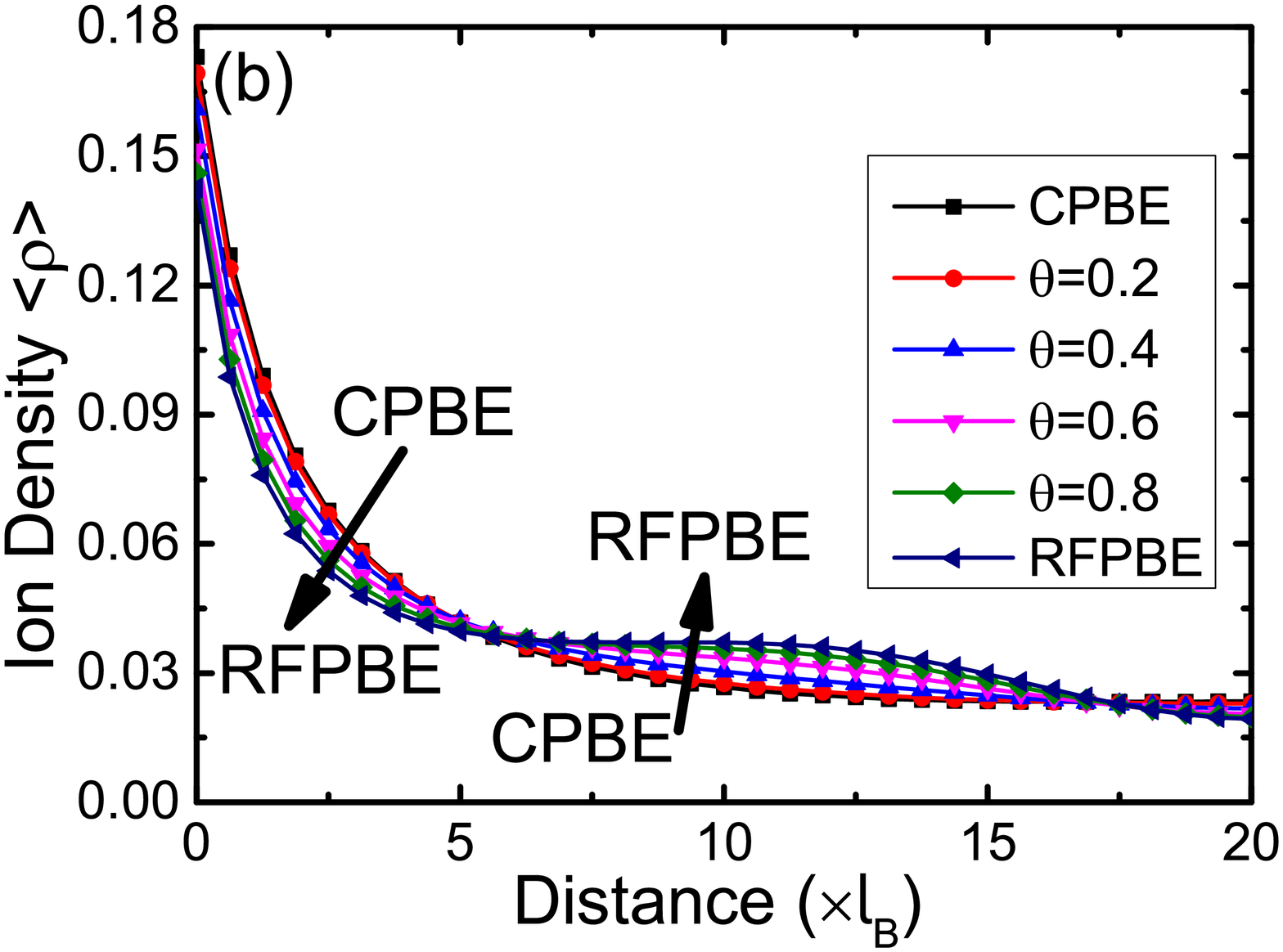}
\includegraphics[width=0.4\textwidth]{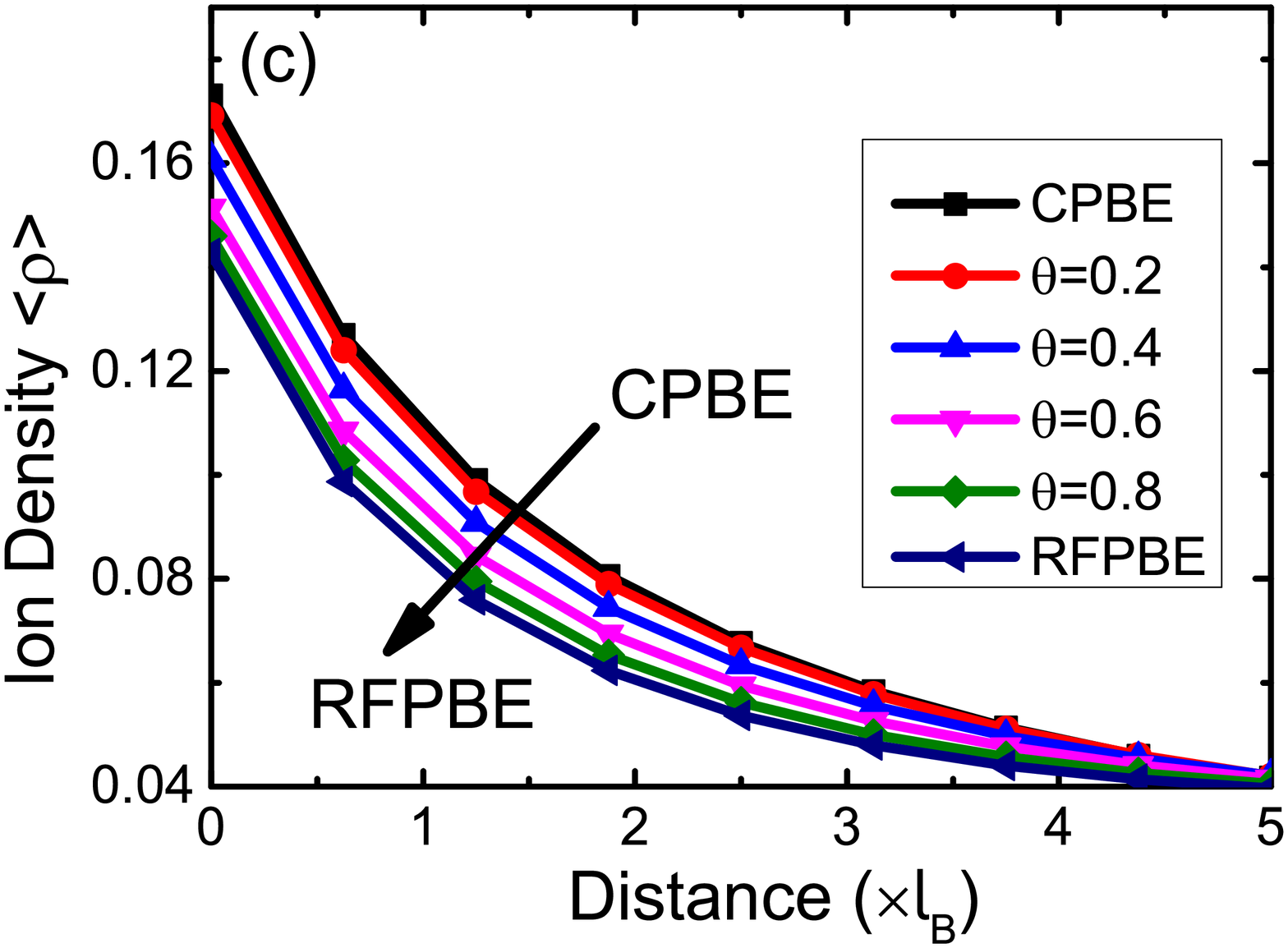}
\includegraphics[width=0.4\textwidth]{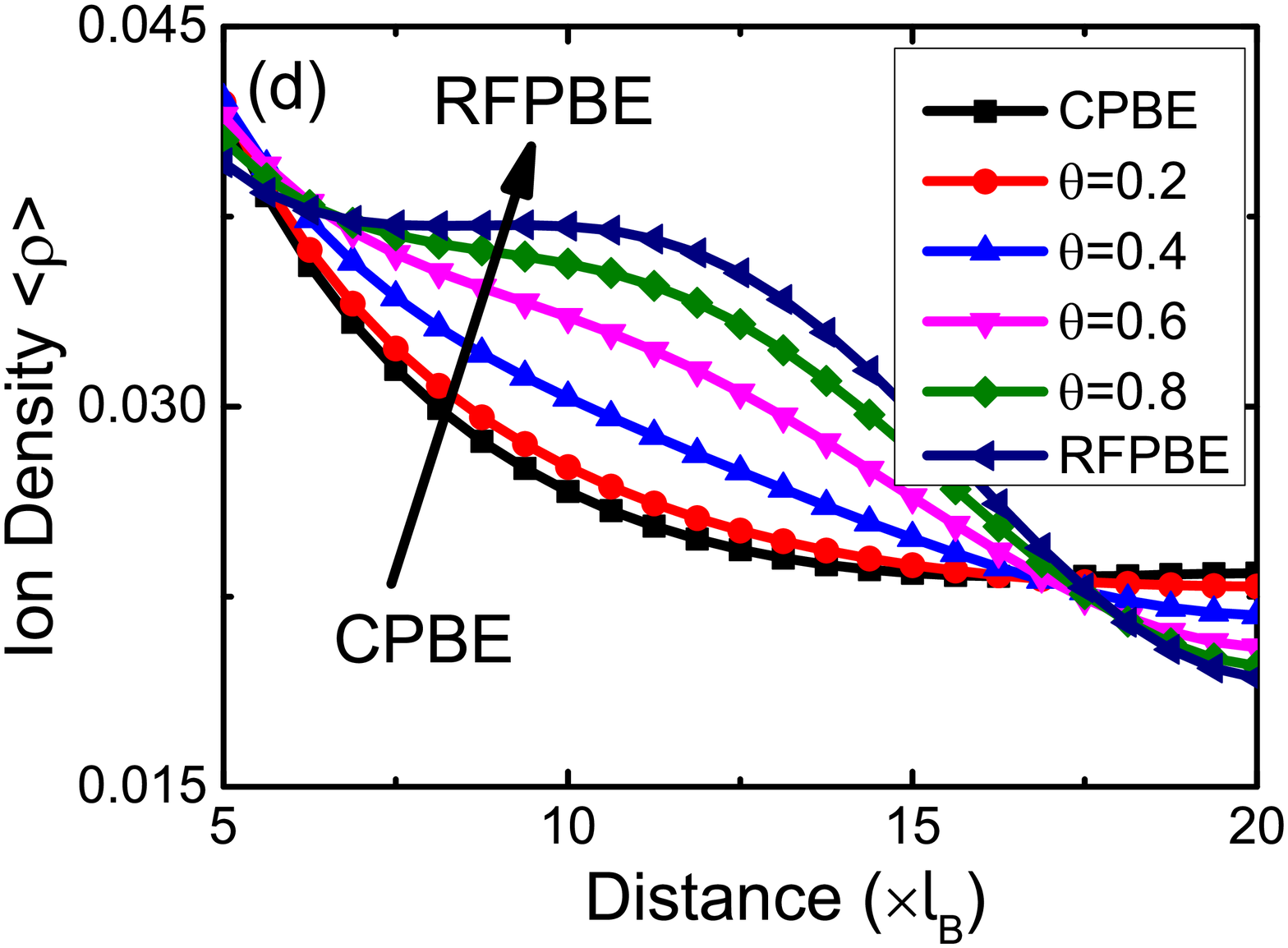}
\caption{Solutions to Eq.(\ref{PIMCspb}) by varying $\vartheta$. Nonlinear effect of the ion fluctuation is enhanced with the increasing of the intensity of the multiplicative noise which is controlled by $\vartheta$.  (a)Electrostatic potentials. (b)Overview of the ion densities with various $\vartheta$, which are replotted in (c) and (d) with different ranges for clarity.}
\end{figure}
Fig.2(a) is for the potential results. With the increasing of $\vartheta$, the curves deviate from the CPBE result and approach the RFPBE curve, meaning that the nonlinear effect of the ion fluctuation is strengthened. Fig.2(b) is an overview of the ion density for various $\vartheta$. For clarity, we plot the data of Fig.2(b) in Fig.2(c) and Fig.2(d) with different ranges. It is shown in the figures that with the increasing of $\vartheta$ the ion density close to the interfaces decreases and the ion density in the domain increases. Similar to the conclusion of Fig.1, the ions are dispersed away from the interfaces and enter into the domain to increase the entropy of the whole system when $\vartheta$ increases to enhance the nonlinear effects of the ion fluctuation.\\

As discussed above, the ion distribution is determined by the competition between the ion fluctuation and the electrostatic force applied by the interfaces. In our BC, the electrostatic force is exerted by the potential drop between interfaces. It is expected that when the potential drop is increased,  more ions are attracted close to the interface at $x=0$, which blocks the dispersing of ions into the bulk. This scenario is confirmed by Fig.3. 
\begin{figure}[!hbp]
\centering
\includegraphics[width=0.5\textwidth]{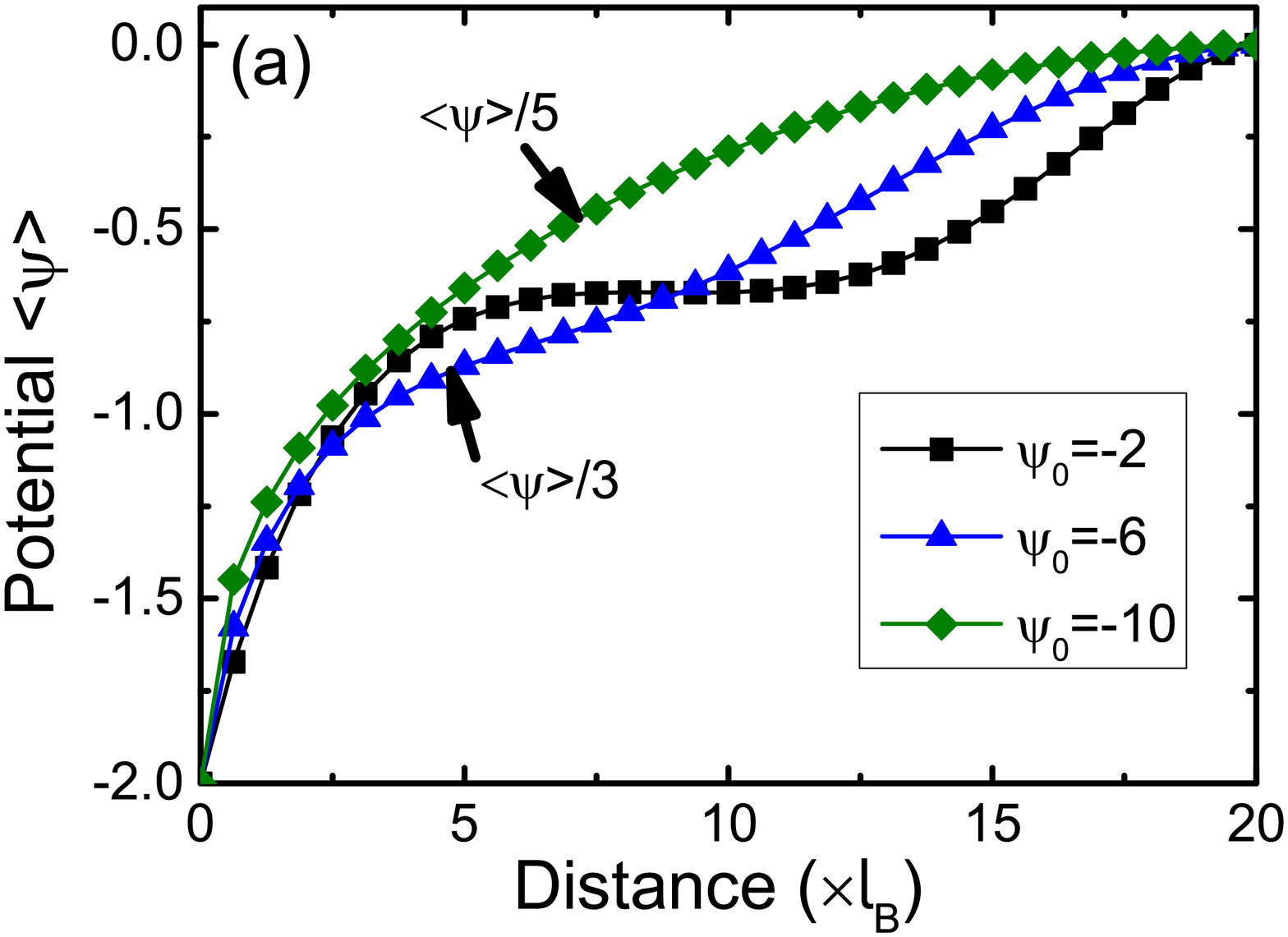}
\includegraphics[width=0.5\textwidth]{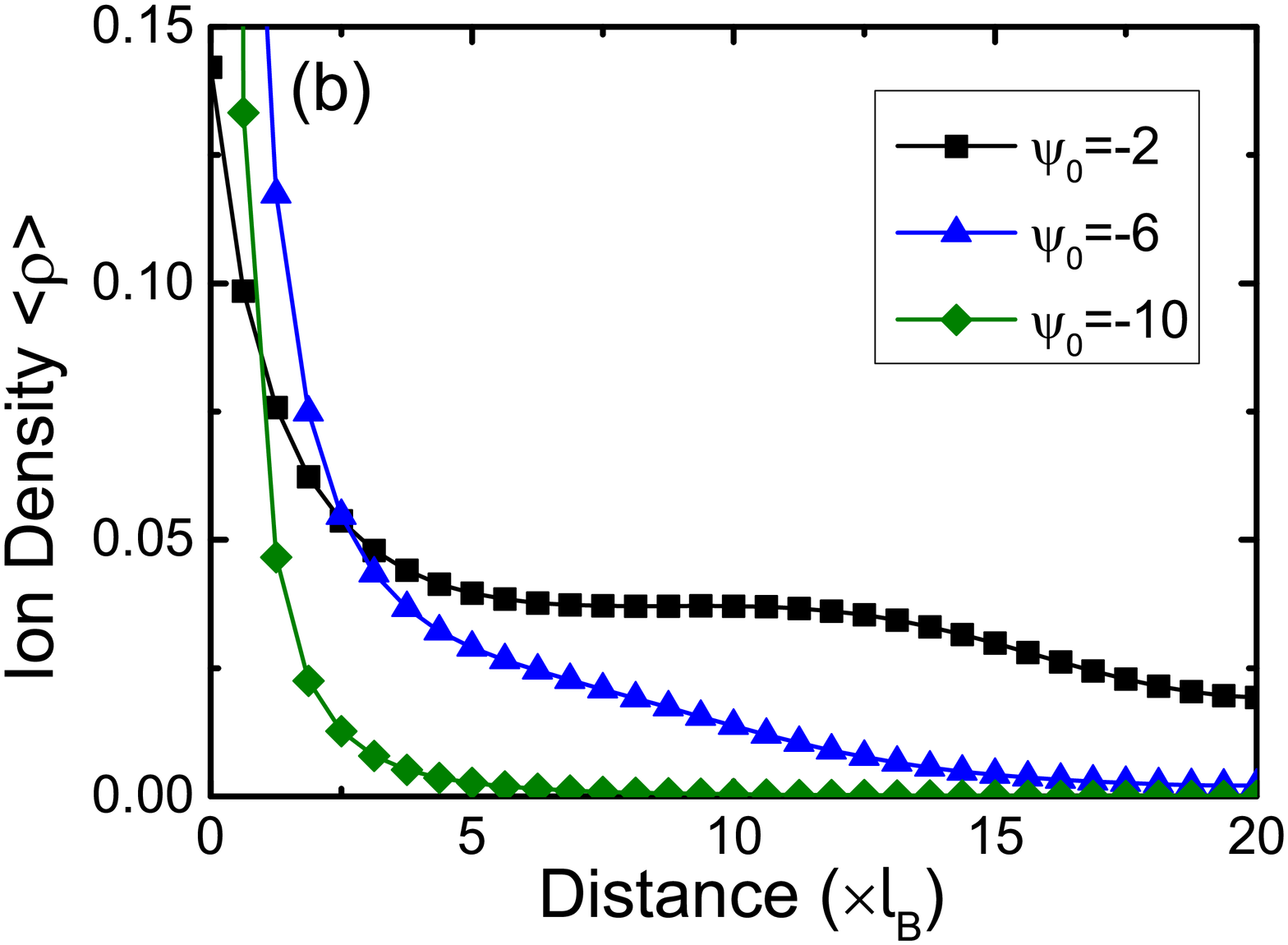}
\caption{Influence of the potential drop on the charged fluid. $\psi_N=0.0$ is fixed while $\psi_0$ is varied to get various potential drops. (a)The electrostatic potentials with various potential drops are scaled in the same plot range for comparison. (b)Ion densities. }
\end{figure}
We fix $M=0.04$, $z=1$ and $\psi_N=0.0$ in this RFPBE calculation, but vary $\psi_0$ to change the potential drop on the charged fluid. For comparison, we normalize all the potential data into the same plot range, shown in Fig.3(a). After the normalization, the three curves have the same start point of $\psi_0=-2$. It shows that the curve of $\psi_0=-10$ behaves different to the other two curves and no such transition segment could be observed in the curve, meaning that the nonlinear effect of the ion fluctuation has been suppressed by the high potential drop. Such phenomena has also been reflected in the ion density shown in Fig.3(b). Since the ion densities close to the interfaces have different scales for the three curves, we plot the main part of the data in Fig.3(b) for comparison. Compared to the other two cases, the high potential drop of $\psi_0=-10$ really attracts more ions to the interface and leaves less ion density in the domain. In this way, the high potential drop of $\psi_0=-10$ dominates over the nonlinear ion fluctuation and no transition segment can be found in the ion density curve. When the potential drop is decreased, the nonlinear effect of the ion fluctuation is strengthened. Ions are dispersed into the domain to increase the entropy of the system, such as in the cases of $\psi_0=-6$ and $\psi_0=-2$.\\

Finally, we study the effect of the ion bulk density $M$ on the ion fluctuation by the RFPBE equation. We set the potential drop as $\psi_0=-2$ and $\psi_N=0$, and vary $M$. $z=1$ is still fixed. When the ion bulk density in the system increases, the probability for the ions to interact with each other is enhanced. Therefore, it is easier for the ions to distribute in the system uniformly with the larger $M$. This point is revealed in Fig.4. 
\begin{figure}[!hbp]
\centering
\includegraphics[width=0.5\textwidth]{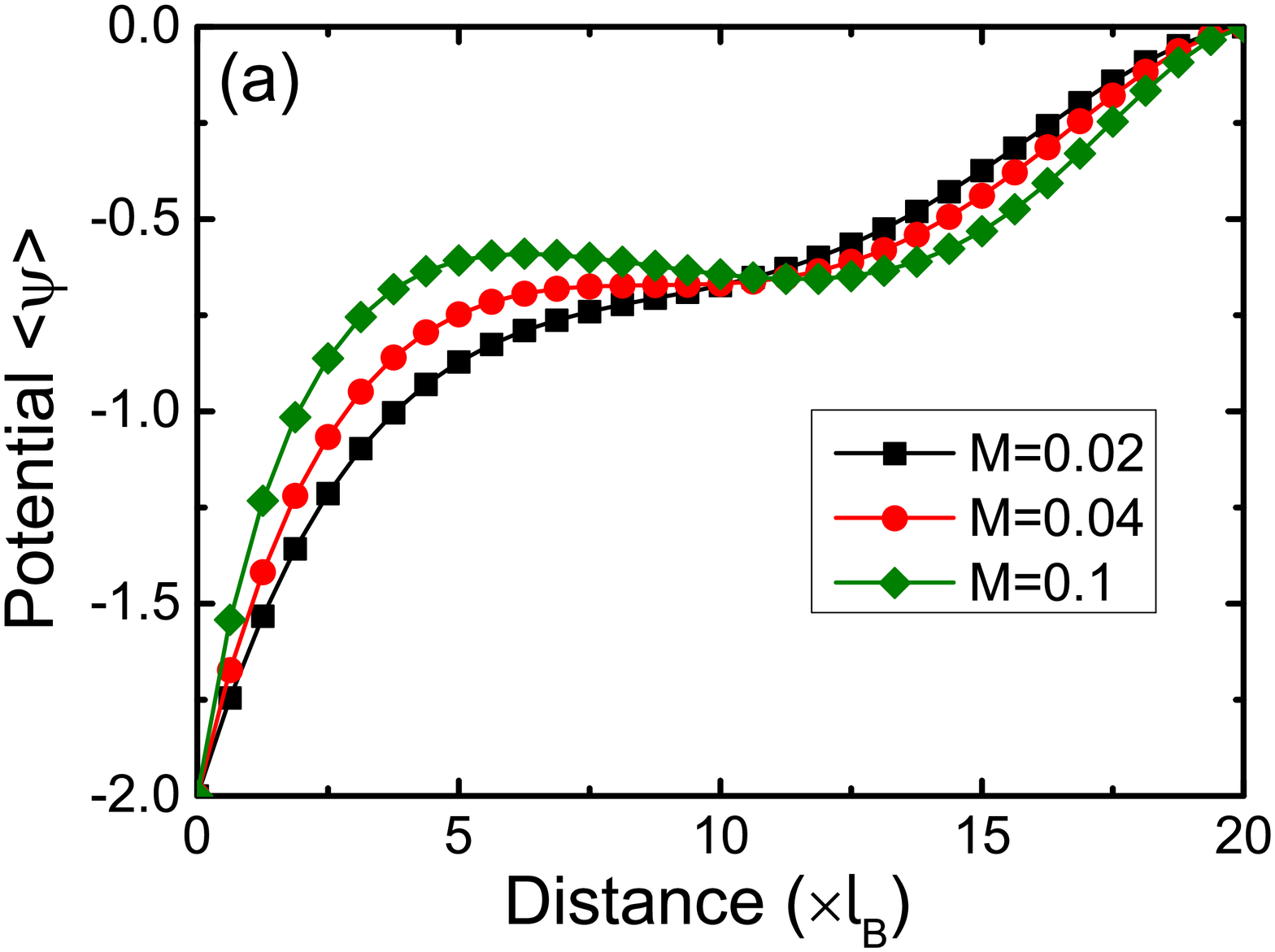}
\includegraphics[width=0.5\textwidth]{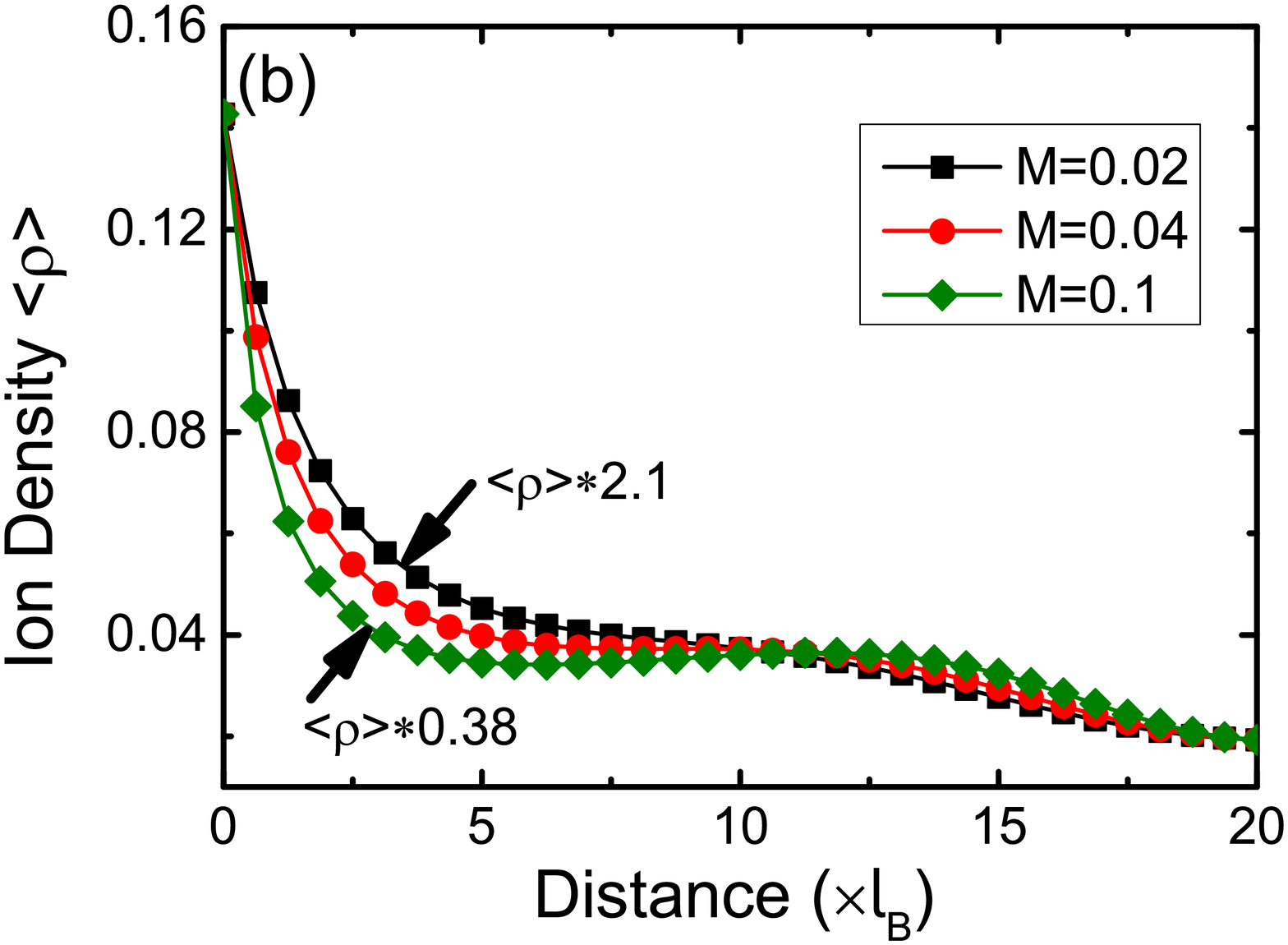}
\caption{Influence of the ion bulk density $M$ on the charged fluid. (a)Electrostatic potentials. (b)Ion densities with various $M$ are scaled in the same plot range for comparison. }
\end{figure}
In the Fig.4(a), the largest ion bulk density $M=0.1$ has the smallest potential gradient at the transition segment in the three cases, showing that the ions distribute much more uniformly in the domain for the larger ion bulk density. Fig.4(b) is for the distribution of the ion density. We have scaled the data in the same plot range for comparison. Fig.4(b) shows that the if the ion bulk density is increased  the ion fluctuation is strengthened and  the electrostatic force is weakened as well. Thus, the nonlinear effect of higher ion bulk density leads to a much more uniform distribution of ions in the higher ion density case, such as in the case of $M=0.1$.\\

In this study, the dielectric constant in the system is homogeneous. For the inhomogeneous systems, we need to implement the gradient of the dielectric into the RFPBE and the PIMC should also be modified, which is still under our study. On this topic, many approaches have been developed to simplify the numerical calculations in solving the complicated self-consistent equations for the inhomogeneous system~\cite{1Hatlo,1Kanduc,2Buyukdagli1}.

\section{Extensions to higher dimensions} 

The RFPBE presented in this paper is generally applicable to any spatial dimensions. In the previous section, we developed the PIMC algorithm and apply it to the one dimensional system. In this section, we extend the PIMC algorithm to higher dimensional systems and apply it to a two dimensional structure. 

\subsection{PIMC for high dimensional systems}
Now, we present the PIMC for high dimensional systems. The general RFPBE reads
\begin{align}
-\nabla^2 \psi=f(\psi)-g(\psi)\eta.
\end{align}
We introduce a vector function $\vec{K}(x)=\frac{1}{g}\nabla \psi$ to transform the above equation to two first order differential equations
\begin{align}
\label{equs2D}
&\nabla \cdot \vec{K}+\vec{K}\cdot \vec{K} \frac{dg}{d\psi}+\frac{f(\psi)}{g(\psi)}-\eta=0, \nonumber \\
&\nabla \psi=g\vec{K}.
\end{align}
For simple notation, we use $h$ to represent $\left( \vec{K}\cdot \vec{K} \right) \frac{dg}{d\psi}+\frac{f(\psi)}{g(\psi)}$ and discretize the system into a lattice. We still use $n$ to index the lattice number and introduce two parameters $\delta W$ and $\delta Q$ by $\delta W=\eta\Delta$ and $\delta Q=(\nabla \cdot \vec{K})\Delta$. Here, we have $\delta W=W_n-W_{n-1}$ and $\delta Q=Q_n-Q_{n-1}$. Then, we discretize the first equation of Eq.(\ref{equs2D}) in Stratonovich sense as
\begin{align}
\label{2Ddiscrete}
Q_{n}-Q_{n-1}+\frac{(h_{n}+h_{n-1})\Delta}{2}=W_{n}-W_{n-1}.
\end{align}
The Jacobian determinant for the variable transformation of Eq.(\ref{2Ddiscrete}) reads
\begin{align}
\frac{dW_{n}}{d Q_{n}} \approx e^{\left(\vec{K}_{n}\cdot \frac{d \vec{K}_{n}}{dQ_{n}}\right)\frac{d g_{n}}{d \psi_{n}}\Delta}.
\end{align} 
The probability for a distribution of the electrostatic potential in the system is
\begin{align}
\label{PI2D}
P(\psi)&=\int \mathcal{D}[\psi K Q]~\delta(\psi-\psi_{B})~\delta(\nabla \psi-g\vec{K})\nonumber \\
& ~~~~\cdot \delta (\nabla \cdot \vec{K}-\delta Q/\Delta) e^{L}.
\end{align}
Here, $\mathcal{D}[\psi K Q]=\prod _n d \psi_n  d \vec{K}_n  \frac{dQ_n}{\sqrt{2\pi \Delta}}$ and $\psi_B$ is the potential applied on the interfaces as the Dirichlet boundary condition. The action $L$ can be split into two components by $L=L_1+L_2$ with 
\begin{align}
\label{action2D1}
L_1&=\int dv \left[ \left( \vec{K}\cdot \frac{d \vec{K}}{dQ} \right) \frac{d g}{d\psi} \right],\nonumber \\
L_2&=\int dv \left\{ -\frac{1}{2} \left[ \nabla \cdot \vec{K}+\vec{K}\cdot \vec{K} \frac{dg}{d\psi}+\frac{f(\psi)}{g(\psi)} \right]^2 \right\}.
\end{align}
$L_1$ is from the Jacobian determination for the variable transformation while $L_2$ is from the Wiener process. $L_1$ can be simplified further into $\sum_{y,z'}\int dx K_x \frac{d g}{d\psi}+\sum_{x,z'}\int dy K_y\frac{d g}{d\psi}+\sum_{x,y}\int dz' K_{z'}\frac{d g}{d\psi}$ for three dimensional (3D) system and $\sum_{y}\int dx K_x\frac{d g}{d\psi}+\sum_{x}\int dy K_y\frac{d g}{d\psi}$ for two dimensional (2D) system. For the one dimensional (1D) system, $L_1$ is simplified into $\int dx K_x\frac{d g}{d\psi}$, which is exactly the same as the corresponding component in Eq.(\ref{action}). Here, $x$,$y$ and $z'$ are coordinates in the systems while $K_x$, $K_y$ and $K_{z'}$ are the components of $\vec{K}$ along $x$,$y$ and $z'$ axes, respectively.\\

In the following study, we take the 2D system as an example. We discretize the system into a lattice and use integers $(n,m)$ to index the lattice sites. $n$ is for $y$ direction while $m$ is for $x$ direction. Correspondingly, the action becomes
\begin{align}
&L_1=\sum _{n,m} \left[K_x^{(n,m)}\Delta_x+K_x^{(n,m)}\Delta_y \right] \frac{d g^{(n,m)}}{d\psi^{(n,m)}},   \nonumber \\
&L_2=-\sum _{n,m} \frac{\Delta}{2} \Big\{ \frac{K_x^{(n,m+1)}-K_x^{(n,m)}}{\Delta_x}+\frac{K_y^{(n+1,m)}-K_y^{(n,m)}}{\Delta_y}\nonumber \\
&+\left[ \left( K_x^{(n,m)} \right)^2+ \left(K_y^{(n,m)} \right)^2 \right] \frac{dg^{(n,m)}}{d\psi^{(n,m)}}+\frac{f^{(n,m)}}{g^{(n,m)}} \Big\}^2.
\end{align}    
Here, $\Delta_x$ and $\Delta_y$ are the mesh lengths along $x$ and $y$ axes, respectively. We have $\Delta=\Delta_x\Delta_y$. Then we use $e^{L_1+L_2}$ for the Metropolis algorithm as we have done for the 1D system in the last section.\\

For the 2D system, we will study the ion fluctuation of charged fluid confined in a rectangular structure, shown in Fig.5(a). The width and the length of the rectangular structure are denoted by $a$ and $b$ respectively. Four potentials applied on the four corners of the structure are indicated in Fig.5(a) as the Dirichlet boundary condition. Along each interface, the potential changes linearly from one corner to the other. In such 2D structure, we still consider only one ionic species and the ions are positive. We have the explicit form for the RFPBE as
\begin{align}
\label{PIMCspb2D}
- \left( \frac{d^2 \psi}{dx^2}+\frac{d^2 \psi}{dy^2} \right)=\frac{zabMe^{-z\psi}}{\int dx dy e^{-z\psi}}-\vartheta \sqrt{ \frac{z^2abMe^{-z\psi}}{\int dx dy e^{-z\psi}}}~\eta.
\end{align}  
Similarly, we define $f(\psi)=\frac{zabMe^{-z\psi}}{\int dx dy e^{-z\psi}}$ and $g(\psi)=\vartheta \sqrt{ \frac{z^2abMe^{-z\psi}}{\int dx dy e^{-z\psi}}}$ for the PIMC calculation. In Eq.(\ref{PIMCspb2D}), $z$  is the charge value of each ion and $M$ is the bulk density of the ions, which have the same meaning as in Eq.(\ref{SPB1}). The charge conservation has also been implemented into Eq.(\ref{PIMCspb2D}). 
\begin{figure}[!tbp]
\centering
\includegraphics[width=0.37\textwidth]{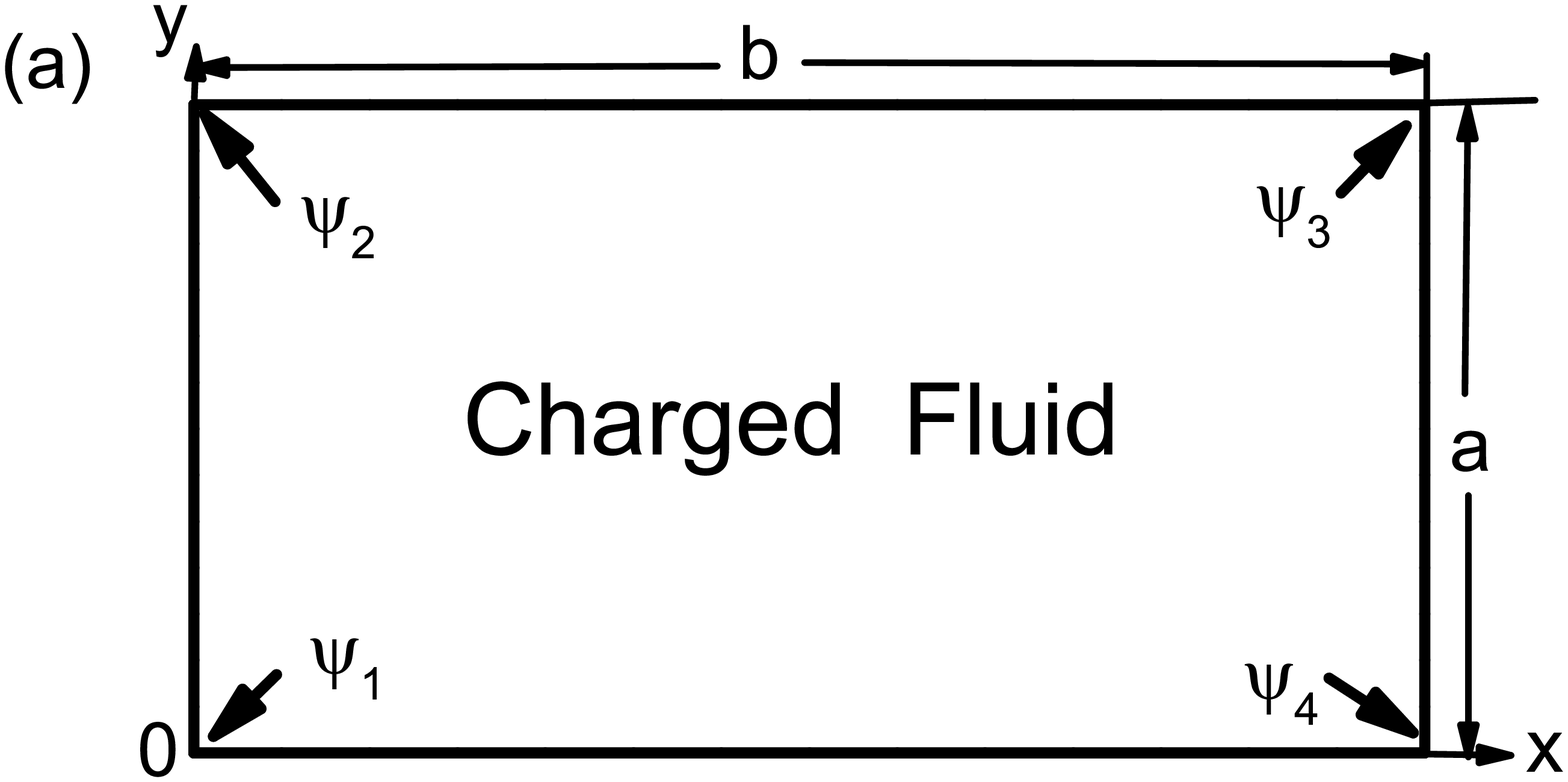}
\includegraphics[width=0.3\textwidth]{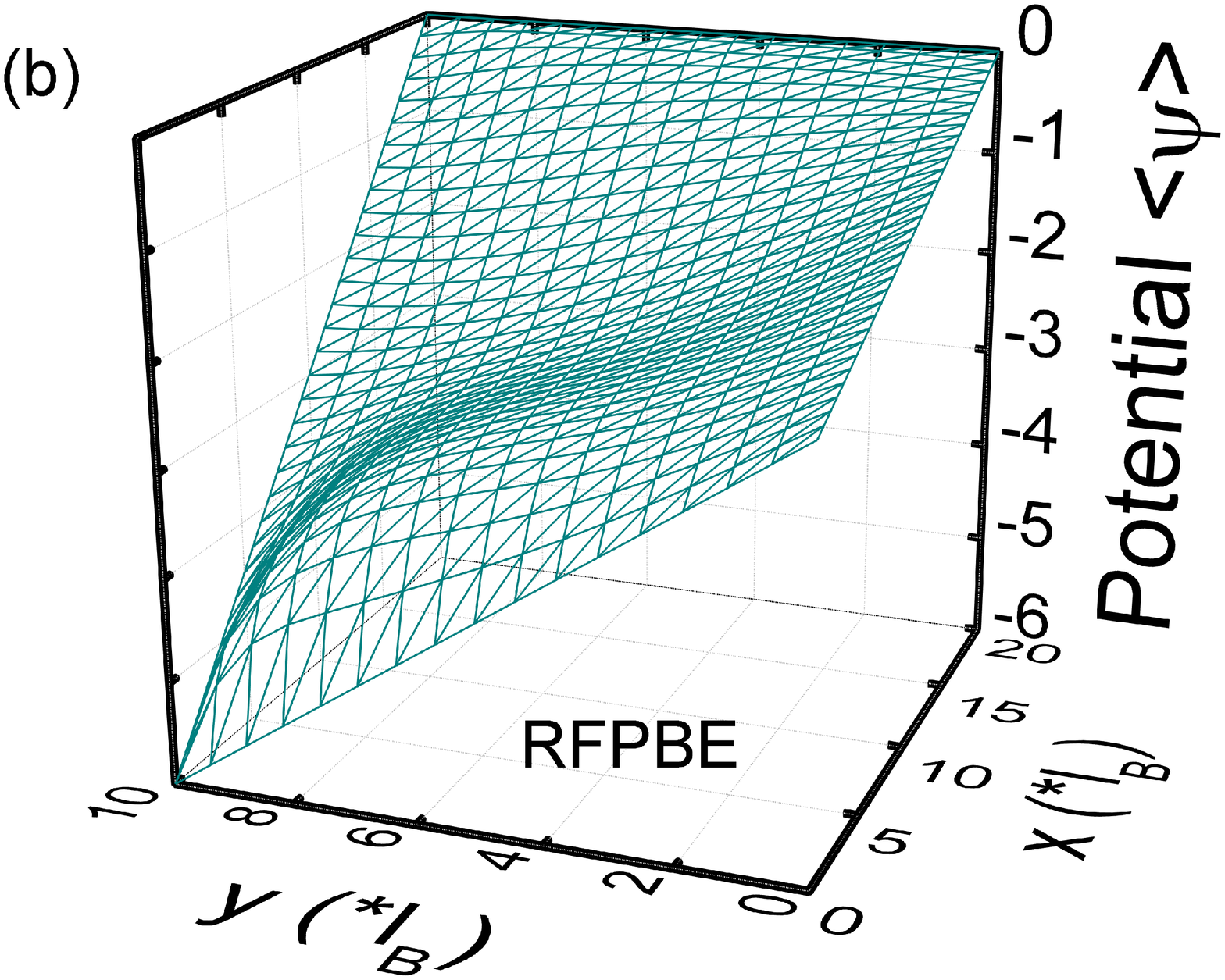}
\includegraphics[width=0.3\textwidth]{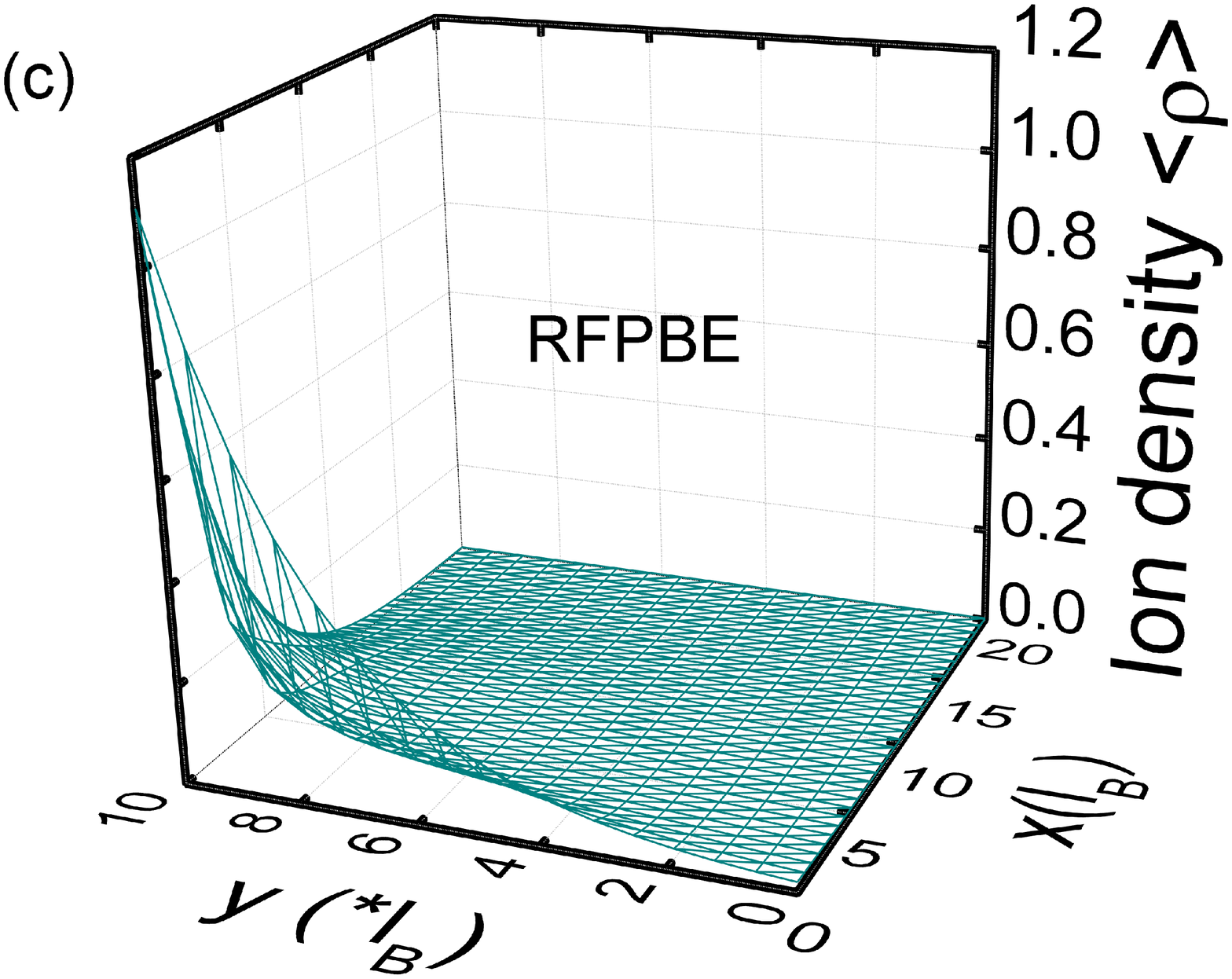}
\includegraphics[width=0.3\textwidth]{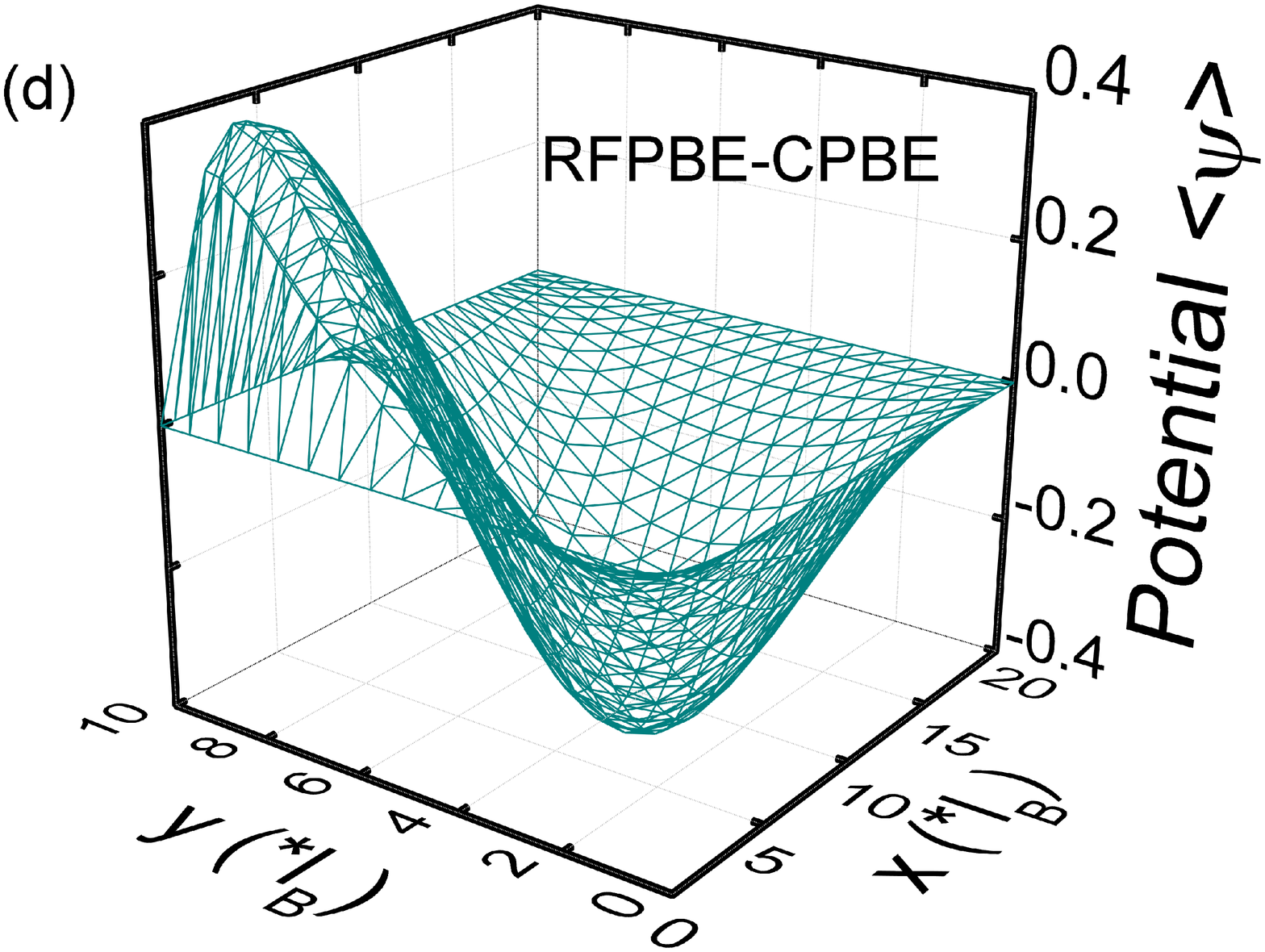}
\includegraphics[width=0.3\textwidth]{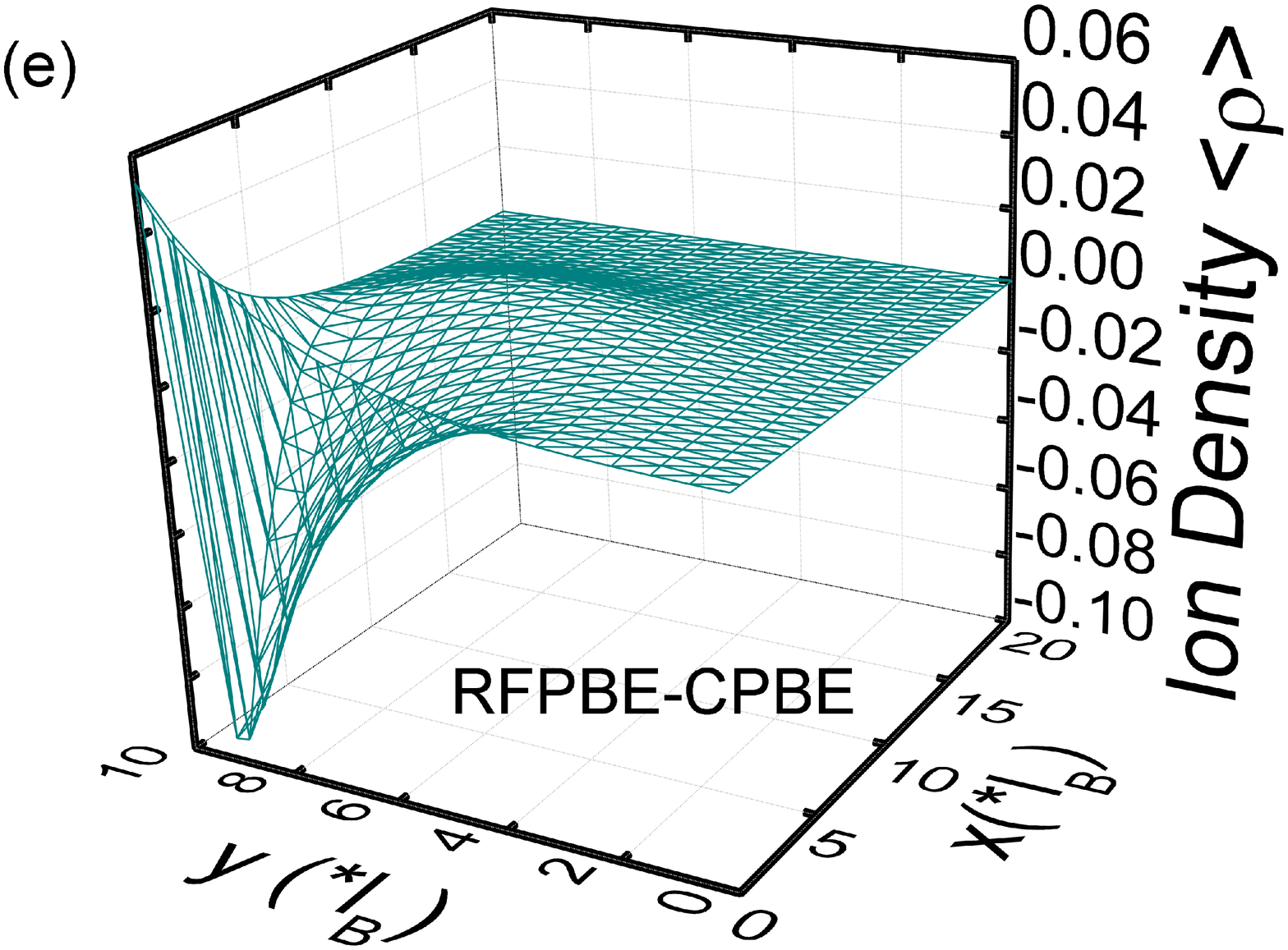}
\caption{(a) Geometry of the 2D structure studied in this paper; (b) and (c):  RFPBE results obtained from  Eq.(\ref{PIMCspb2D}) with $\vartheta=1.0$; (d) and (e):  RFPBE results minus those of CPBE which is obtained from Eq.(\ref{PIMCspb2D}) with $\vartheta=0.01$.}
\end{figure} 
 
 \subsection {Results for 2D system}
In Fig.5, we present the results obtained from Eq.(\ref{PIMCspb2D}) by using the PIMC method for the 2D structure. $\vartheta=1$ in Eq.(\ref{PIMCspb2D}) is for the  RFPBE and $\vartheta=0.01$ is for the CPBE, as we have done for the 1D system. We use the following parameter values: $a=10.0$, $b=20.0$, $\psi_1=-2.0$, $\psi_2=-6.0$, and $\psi_3=\psi_4=0.0$. $M=0.04$ and $z=1$ are still used for illustration. \\

Fig.5(b) shows the potential distribution for the 2D system. The potentials at the four corners are fixed and the details of the potential distribution in the charged fluid have been well captured by our PIMC method. At the position of $x=0$ and $y=10$, the potential $\psi_2=-6.0$ is the lowest, which attracts almost all the positive ions to this position. Such phenomena has been exhibited in Fig.5(c). In order to show the effects of the ion fluctuation in the charged fluid clearly, we subtract the potential of CPBE from that of  RFPBE and present the potential difference in Fig.5(d). It shows that the  RFPBE potential is larger than the CPBE potential close to the position of $\psi_2$. But in the center of the system, the RFPBE potential is lower than the CPBE potential. Since lower potential attracts more positive ions, that means more positive ions are expelled from the boundaries and concentrate in the bulk of the system, which is consistent with the result of Fig.5(e). In Fig.5(e), we subtract the ion density of CPBE from that of RFPBE and present the difference of the ion density. It shows that the ion density of the  RFPBE is less than that of the CPBE close to the position of $\psi_2$, but larger in the bulk of the system. The physics behind Fig.5 is the same as we have discussed for the 1D system. That is, the ion fluctuation increases the entropy of the whole system and has the effect of distributing the ion density uniformly. 
  
\section{Conclusion}
In the charged fluid, the ion distribution is determined by the competition between the ion fluctuation and the electrostatic force applied by the boundaries. The CPBE has been applied to study the ion distribution. It averages the ion fluctuation and misses the nonlinear effects of the ion fluctuation, which is not enough to describe the ion distribution precisely. Based on the field theory, we derive the RFPBE with a multiplicative noise added into the CPBE.  The multiplicative noise captures the nonlinear effect of the ion fluctuation. To solve the RFPBE, we develop a path integral Monte Carlo method to sample the potential paths.\\

Numerical results for one and two dimensional systems show that the nonlinear ion fluctuation strengthens the dispersion of ions away from the interfaces into the domain to increase the entropy of the charged fluid. Increase of the electrostatic force by the boundaries enhances the attraction of ions to the interfaces and weakens the nonlinear effect of the ion fluctuation. On the other side, increase of the ion bulk density enhances the nonlinear ion fluctuation and weakens the electrostatic forces by the boundaries.\\

The RFPBE obtained in this work is independent of the coupling between the charged fluid and interfaces, and it focuses on the ion fluctuation of the charged fluid itself. The coupling is taken into account in the solution as the boundary conditions. Thus, the present RFPBE is general and can be applied to various cases. As we have mentioned, the steric effect is missing in the RFPBE. For the charged fluid with a high ion density, we need to implement the steric effect in the equation to capture the proper effect of the ion fluctuation. This issue is still under study.\\ 


\end{document}